\documentclass[12pt,preprint]{aastex}

\slugcomment{to be Submitted to the Astrophysical Journal Supplement Series}

\shorttitle{A {\it FUSE} Catalog of CVs              
}
\shortauthors{Godon et al.}

\begin{document}

\title{
An Online Catalog of Cataclysmic Variable Spectra from the  
{\it Far Ultraviolet Spectroscopic Explorer} 
} 

\author{
Patrick Godon\altaffilmark{1}, Edward M. Sion 
} 
\affil{Astronomy \& Astrophysics, Villanova University, \\ 
800 Lancaster Avenue, Villanova, PA 19085, USA}
\email{
patrick.godon@villanova.edu; edward.sion@villanova.edu } 

\author{Karen Levay} 
\affil{Space Telescope Science Institute, 
Baltimore, MD 21218}
\email{klevay@stsci.edu} 

\author{Albert P. Linnell, Paula Szkody}
\affil{Department of Astronomy, University of Washington, Box 351580, 
Seattle, WA 98195-1580}
\email{ linnell@astro.washington.edu ; szkody@astro.washington.edu }  

\author{Paul E. Barrett}
\affil{United States Naval Observatory, Washington, DC 20392}
\email{ barrett.paul@usno.navy.mil } 

\author{Ivan Hubeny}
\affil{Steward Observatory and Department of Astronomy and Astrophysics,
University of Arizona, Tucson, AZ 85721}
\email{ hubeny@as.arizona.edu } 
 
\and

\author{William P. Blair}
\affil{Henry A. Rowland Department of Physics and Astronomy, 
The Johns Hopkins University,
Baltimore, MD 21218}
\email{wpb@pha.jhu.edu} 

\altaffiltext{1}{Visiting at the Johns Hopkins University,
Henry A. Rowland Department of Physics and Astronomy, 
Baltimore, MD 21218}

\begin{abstract} 

We present an online catalog containing spectra and supporting
information for cataclysmic variables that have been
observed with the {\it Far Ultraviolet Spectroscopic Explorer (FUSE)}.
For each object in the catalog we list some of the basic system parameters
such as (RA,Dec), period, inclination, white dwarf mass, as well as
information on the available {\it FUSE} spectra: data ID, observation date and time,
and exposure time. In addition, we provide parameters needed for the
analysis of the {\it FUSE} spectra such as the reddening E(B-V), distance,
and state (high, low, intermediate) of the system at the time it was observed.  
For some of these spectra we have carried out model fits to the continuum
with synthetic stellar and/or disk spectra using the codes TLUSTY and 
SYNSPEC. We provide the parameters obtained from these model fits; this
includes the white dwarf temperature, gravity, projected rotational velocity 
and elemental abundances of C, Si, S and N, together with the disk mass
accretion rate, the resulting inclination and model-derived distance (when unknown).  
For each object one or more figures are provided (as gif files) with      
line identification and model fit(s) when available. The {\it FUSE} spectra
as well as the synthetic spectra are directly available for download
as ascii tables. References are provided for each object as well as
for the model fits. In this article we present 36 objects, and 
additional ones will be added to the online catalog in the future. 
In addition to cataclysmic variables, we also include a few related objects, 
such as a wind accreting white dwarf, a pre-cataclysmic variable
and some symbiotics. 

\end{abstract} 

\keywords{accretion, accretion disks - novae, cataclysmic variables
- white dwarfs}
 
\section{Introduction}

This article presents the first version of an on-going online 
catalog of Far Ultraviolet Spectroscopic Explorer ({\it FUSE}; 
\citep{moo00})
spectra of cataclysmic variables (CVs).
The catalog includes CV types and subtypes, as follows: 
dwarf novae (U Gem, Z Cam, SS Cyg, WZ Sge and SU UMa subtypes), 
nova-likes (VY Scl/anti-DN, UX UMa, SW Sex subtypes), 
and magnetic systems (IPs, Polars, DQ Her, AM Her subtypes). 
In addition the catalog also includes a few miscellaneous objects related 
to CVs such as novae (of all types),
symbiotics, and pre-cataclysmic variables. 
In Table 1 we list the objects in the catalog as of March 2012. 
For each object the table includes RA (J2000), DEC (J2000), 
type and subtype, binary orbital period, system inclination,
reddening value, white dwarf mass, and distance.   
For each object, there is (i) at least one {\it FUSE} spectrum, 
(ii) possibly a theoretical fit to the spectrum (when the continuum 
can be modeled successfully),
and (iii) basic data about the system including figures 
and downloadable ascii tables.  
The state (outburst/high, quiescence/low or
intermediate) of each system at the time of observation
is specified on the {\it object page} (see sec.2.4).  

While \citet{fro12} has presented a summary of {\it FUSE} spectra
of cataclysmic variables, there are similarities and differences
between their work and our catalog. 
The {\it FUSE} survey of \citet{fro12}
is an atlas of all the existing {\it FUSE} spectra of all CVs, 
which includes basic information about each system (type, orbital
period and inclination) and the {\it FUSE} exposures (data ID, exposure
time, data quality, observation date and whether the data were obtained
during a high/outburst or low/quiescent state).  
The spectra are discussed and characterized as a function of 
CV subtype: DNe in quiescence, which are divided into those that exhibit
a flat continuum with emission lines and those exhibiting 
absorption lines (broad Lyman lines and narrower metal lines); 
DNe in outburst; 
nova-likes subdivided into narrow absorption lines, broad absorption
lines, and with emission lines; intermediate polars; and polars. 
Intrinsic spectral lines as well as prominent ISM lines are tabulated. 
As such the paper of \citet{fro12} is really an Atlas of {\it FUSE} spectra
of CVs, containing 99 objects where the spectra are catalogued according
to their characteristics. The spectra are downloadable in fits format
directly from the website (MAST) where a preview is presented including
basic line identification (which is the same for all the spectra).      
 
In comparison, our catalog is more oriented toward the modeling 
and analysis of the spectra. 
The {\it FUSE} spectra are downloadable as
ascii tables which can then be easily used with computer modeling programs 
often used by theorists and/or compared directly with theoretical
spectra obtained from (e.g.) TLUSTY and SYNSPEC \citep{hub88,hub95}.    
In addition to inclination and orbital period, our catalog also 
includes parameters that  
are usually needed in the modeling of spectra, such as white dwarf
mass,  distance, and reddening (when these are known). 
We include a preview of the spectrum spread over three panels for
better viewing, with individual line identifications for each object
separately, including ISM molecular hydrogen lines.  
When possible we provide results from the spectral modeling, including
downloadable theoretical spectra in ascci format and additional system
parameters such as mass accretion rate, WD temperature, stellar 
rotation rate, and chemical abundances.     
We also provide a list of references for each object.  
At the present time the catalog is still growing,  
and systems are being added with time.  
The catalog will eventually include all CVs and related objects 
that were observed with {\it FUSE}. 
In the near future the catalog will also include a large database 
of theoretical
spectra of (solar composition) WDs covering a wide range of temperatures 
and gravity. 

\section{Processing of the {\it FUSE} Spectra}  

The {\it FUSE} telescope and its performance are
described in detail by \citet{moo00,sah00},
and we give here only a short overview of some of its features
and characteristics, as needed to explain and justify some of the 
procedures we followed to extract the final spectra.  
Most of the {\it FUSE} data were obtained 
through the 30"x30" LWRS Large Square Aperture in TIME TAG mode,  
and were processed with the latest and final version of 
CalFUSE (v3.2.3; \citet{dix07}).   
The {\it FUSE} data come in the form of eight spectral segments 
(SiC1a, SiC1b, SiC2a, SiC2b, LiF1a, LiF1b, LiF2a, and LiF2b) which
are combined together to give the final {\it FUSE} spectrum.   
The spectral regions covered by the
spectral channels overlap, and these overlap regions are then used to
renormalize the spectra in the SiC1, LiF2, and SiC2 channels to the flux in
the LiF1 channel. We then produce a final spectrum that covers the
full {\it FUSE} wavelength range 905-1187~\AA . 

The low sensitivity 
portions of each channel are usually discarded because of the low S/N there.
In most channels there exists a narrow dark stripe of decreased flux
in the spectra running in the dispersion direction, 
known as the ``worm'', which can attenuate as much as
50\% of the incident light in the affected portions of the
spectrum; - this is due to shadows thrown by the repeller grid wires 
above the detector.  
Because of the temporal changes in the strength and position of the 
``worm'', CalFUSE cannot correct target fluxes for its presence. 
Therefore, we carried out a visual inspection of the {\it FUSE} channels to
locate the worm and we manually discarded that  
portion of the spectrum.
CalFUSE, for the most part, takes care of removing several additional
well-known issues, such as {\it jitter} and {\it event bursts}. For faint
sources {\it background subtraction} can be a problem for CalFUSE and we
did have to remediate this problem by either ignoring or discarding
those portions of the spectra affected by bad background subtraction.     

Due to these problems, for some of the systems, we had to discard 
the two SiC channels completely. 
For other systems, large portions of the SiC channels had to 
be removed and, as a consequence, some spectra exhibit a
gap around 1080~\AA .
The channels that were the most reliable were the 1aLiF, 
1bLiF and 2aSiC. The 2aLiF channel was moderately
reliable and was used especially when the worm was affecting
the 1bLiF channel. The 1aSiC and 1bSiC channels were the most  
discarded, together with the 2bLiF channel. 
The 2bSiC channel was discarded only when it was found to
be of much lower quality than the LiF channels.  

In the final step, we combined the individual exposures 
(unless otherwise specified) and channels to create a
time-averaged spectrum weighting
the flux by the (good) exposure time and sensitivity of the
input exposure and channel of origin, using a suite of FORTRAN programs, 
unix scripts and IRAF procedures (which were all writen specifically for this
purpose). 

The observation log for the initial objects included in the catalog 
is given in Table 2, where we list the 
{\it FUSE} data ID, the observation date, the observation time,
the good exposure time and the state in which the system was observed
(i.e. high, low or intermediate). 

Details of the data processing for each individual system are provided in
the reference(s) given in the link below the figure(s).
Some of these details are also listed individually for 
each object in the ``Description Panel'' (see appendix B). 

\section{The {\it FUSE} Spectral Lines}  

The main emitting components contributing to the {\it FUSE} spectra of CVs 
are the accretion disk and the WD. Usually, the disk dominates
during the high state (or outburst), while the WD dominates during 
the low state (or quiescence). The main feature
of the  spectra is the broad Ly$\beta$ absorption feature, which can
easily be used to assess the gravity and temperature of the emitting
gas (when this feature is not affected by e.g. large rotational broadening,
strong O\,{\sc vi} emission/absorption lines, and so on).  
At higher temperatures,  
as the continuum rises in the shorter wavelengths,
the higher orders of the Lyman series also become visible
and can also be used (when possible) to accurately assess the temperature
and gravity of the emitting gas. 

Additional broad absorption lines of metals (C, S, Si, ..) are detected 
and help determine the chemical abundances and projected rotational
velocity of the emitting gas.   
In the present spectra, the main absorption features observed 
are: 
C\,{\sc ii} (1010~\AA),  
C\,{\sc iii} (1175~\AA),   
Si\,{\sc iii} (1108-1114~\AA\ and 1140-1144~\AA), 
Si\,{\sc iv} (1067~\AA\ and 1120-1130~\AA), 
S\,{\sc iv} (1073~\AA),  
and 
N\,{\sc ii} (1085~\AA\ when not contaminated by air glow).   

On top of the spectrum, broad emission lines are also found in some of
the systems,
mainly the  
O\,{\sc vi} doublet and 
C\,{\sc iii} (977~\AA\ and 1175~\AA), indicative of wind activity. 

In Table 3 we provide a list of the ions and 
spectral lines expected for these sources
in the {\it FUSE} spectral range.  
The wavelengths are the theoretical rest wavelengths in \AA\ . 

For al the systems we assign four line identification quality values 
as follows. \\ 
A - a fully reliable line identification without contamination, the central wavelength
of the line can easily be determined. \\
B - a reliable line identification with possible contamination or/and blended with another line, 
the central wavelength of the line cannot be determined with precision. \\
C - a possible line identification, but less reliable than B, due to low S/N
and other factors. A spectral feature is seen but is not identified with certainty. \\
D - the line is either not detected, or the detection is not reliable at all; the line identification
on the figure is tentative or is only informative (to better show that the line is missing
where it should be). A very noisy spectral feature is seen, or no spectral feature is
detected at all (flat continuum). \\ 
For all the systems, 
in Table 4 we tabulate these quality values for the most common spectral lines seen in CVs. 

Most of the {\it FUSE} spectra 
show some ISM molecular hydrogen absorption. The most affected
targets reveal a spectrum literally ``sliced'' 
at almost equal intervals ($\sim$12~\AA);  
starting at wavelengths around 1110~\AA\ and continuing
towards shorter wavelengths all the way down to the hydrogen cut-off around
915~\AA. In the affected {\it FUSE} spectra, we identified the 
most prominent molecular hydrogen absorption lines by their band
(Werner or Lyman), upper vibrational level (1-16), and rotational transition
(R, P, or Q) with lower rotational state (J=1,2,3). 
The most common ISM lines are tabulated in Table 5.  

In addition,    the targets that are weak 
{\it FUSE} sources exhibit sharp emission lines from air 
glow (geo- and helio-coronal in origin; some of which are due to sunlight
directly reflected inside the telescope),  such as 
the H\,{\sc  i} series, 
S\,{\sc  vi} (934~\AA\  and 944~\AA),  
O\,{\sc  vi} doublet, 
C\,{\sc  iii} (977~\AA), and 
He\,{\sc  i} (1168~\AA).  

For each system, we mark on the (online) figures the lines that are detected
as well as lines that are commonly seen for comparison.

\section{Spectral Modeling} 
 
\subsection{The Synthetic Stellar Spectral Codes}  

We created model spectra for
high-gravity stellar atmospheres using the codes 
TLUSTY and SYNSPEC
(http://nova.astro.umd.edu ; \citet{hub88,hub95}).

Atmospheric structure is computed (using TLUSTY) assuming either a LTE 
or a NLTE atmosphere for a given stellar surface temperature and gravity,  
where the switch to NLTE is made for hot models ($T>30,000~K$).
In that manner, we generate photospheric models with effective
temperatures ranging from 12,000~K to 75,000~K in increments of 
about 10\% (e.g. 1,000~K for T$\sim$15,000~K and 5,000~K for 
T$\sim$70,000~K).
We choose values of the surface gravity $log(g)$ ranging 
between 7.0 and 9.5. 
The (converged) output stellar atmosphere model is then used
as an input for the synthetic spectrum generator code SYNSPEC,
and finally the code ROTIN is used to simulate the broadening
of the lines due to stellar rotation and/or instrumentation. 
Limb darkening is also included in ROTIN.  
The stellar rotational velocity $V_{rot} \sin(i)$ varies 
from a few km~$s^{-1}$ to 1000~km~$s^{-1}$.
In order to fit the absorption features of the spectrum, we also vary
the chemical abundances of the main species C, N, S and Si.
For any WD 
mass, there is a corresponding radius, or equivalently, one single value of
$log(g)$ \citep{ham61}, which also depends on the temperature
and composition of the WD \citep{woo95,pan00}. 

The same suite of codes can also be used to generate spectra of
accretion disks \citep{wad98}  
based on the standard alpha disk model \citep{sha73}. 
The standard disk model assumes a geometrically thin axi-symmetric
disk, one-dimensional, with a nearly Keplerian profile
in which the viscosity is due to 
MHD turbulence. The density and temperature
profiles are function of $r$, and the energy dissipated
due to the shear between adjacent rings of matter 
is radiated locally in the vertical direction ($z$).
The disk model consists in dividing the disk 
into N rings of radii $r_i$ $(i=1,2,..N)$,
where the temperature $T(r_i)$ and density $\rho(r_i)$ of each ring  
are given by the standard disk model.  
The TLUSTY \& SYNSPEC codes are then used to generate spectra
for the disk rings, given the white dwarf mass, mass accretion
rate, and radius of the ring. The spectra of the rings are then
added together to generate a disk spectrum for a given inclination
and assuming limb darkening.   
In the present work we use disk spectra from the grid of 
solar composition disk spectra generated by 
\citet{wad98} (assuming LTE only)  
as well as disk spectra that we generated using both the
LTE and NLTE options for disk rings, as well as in some
cases varying the chemical abundances. A detailed description
of the procedure to generate such disk spectra is given 
by \citet{wad98}.  
Composite models of WD+disk take into account the geometry
of the system and account for the visibility of the WD. 

\subsection{Modeling the ISM Hydrogen Absorption Lines}   

For most systems showing ISM atomic and molecular hydrogen 
absorption lines, we identify these lines in the figures to
avoid confusing them with the WD lines. For some systems, however,
ISM lines are deep and broad 
and some of the WD lines (such as 
S\,{\sc iv} (1062.6~\AA\ \& 1073~\AA\ )  are located 
at almost the same wavelengths as some of the ISM lines.   
For that reason, we model the ISM lines; this helps us 
not only to differentiate between the WD lines and the ISM lines, 
but it improves the overall spectral fit,  and provides us with 
the atomic and molecular column densities. 

The ISM spectra models are generated using a program developed by 
one of us (PEB).  
This program uses a custom spectral fitting package to estimate the temperature
and density of the interstellar absorption lines of atomic and
molecular hydrogen.  The ISM model assumes that the temperature, bulk
velocity, and turbulent velocity of the medium are the same for all
atomic and molecular species, whereas the densities of atomic and
molecular hydrogen, and the ratios of deuterium to hydrogen and metals
(including helium) to hydrogen can be adjusted independently. The
model uses atomic data of \citet{mor00,mor03}  
and molecular data of \citet{abg00}. 
The optical depth calculations of molecular
hydrogen have been checked against those of \citet{mor03}.   
The ratios of metals to hydrogen and deuterium to
hydrogen are fixed at $0$ and $2 \times 10^{-5}$, respectively,
because of the low
signal-to-noise ratio data.  The wings of the atomic lines are used to
estimate the density of atomic hydrogen and the depth of the
unsaturated molecular lines for molecular hydrogen.  The temperature
and turbulent velocity of the medium are primarily determined from the
lines of molecular hydrogen when the ISM temperatures are $<$ 250K.

The ISM absorption features are best  modeled and displayed when the
theoretical ISM model (transmission values) is combined with a
synthetic spectrum for the object (namely a WD synthetic spectrum,
a disk spectrum or a combination of both).

\subsection{Synthetic Spectral Model Fitting}  

Before carrying out a synthetic spectral fit,
we masked portions of the spectra with strong emission lines,
strong ISM molecular absorption lines, detector noise and air glow.
These regions of the spectra are somewhat different for each object and 
are not included in the fitting. The regions excluded from the fit
are in blue in the figures.
The excluded ISM quasi-molecular absorption lines are 
marked with vertical labels in the figures.

After having generated grids of models for each target, 
we use FIT, a $\chi^2$ minimization routine (see e.g. \citet{pre92}), 
to compute the reduced $\chi^2_{\nu}$   
($\chi^2$   per number of degrees of freedom ${\nu}$) 
and scale factor values for each model fit.  
While we use a $\chi^2$ minimization technique, we do not 
blindly select the least $\chi^2$ models, but we examine the models 
that best fit some of the features such as absorption
lines 
and, when possible, the slope of the wings of the broad Lyman
absorption features. When possible, we 
also select the models that are in agreement
with the distance of the system (or its best estimate).  

The flux level at 1000~\AA\  (between Ly$\beta$  and Ly$\gamma$) 
is close to zero for temperatures below 18,000~K; at 30,000~K it is about
50\% of the continuum level at 1100~\AA\
and it reaches 100\% for $T>45,000$~K.  
At higher temperature ($T>50,000$~K) the spectrum becomes fairly flat 
and there is not much difference in the shape of the spectrum 
between (say) a 50,000~K and a 80,000~K model.  
When fitting the shape of the spectrum in such a manner, 
an accuracy of about 500-1,000~K is obtained, due to the S/N.
In theory, a fine tuning of the temperature (say to an accuracy of about
50~K) can be carried
out by fitting the flux levels such that the distance to the system
(if known) is matched. However, the fitting to the
distance depends strongly on the radius (and therefore the mass)
of the WD. In most of the systems the mass
of the WD and the distance are 
unknown and therefore the temperature of a system is often accurate
to only $\sim 500-1000$~K.  
Furthermore, since the Ly$\beta$ profile  depends on both
the temperature and gravity of the WD, the accuracy of the
solution is sometimes further decreased as there is 
a small degeneracy in the solution,
namely the solution spreads over a small region of the $\log(g)$ and $T$ 
parameter space. And last, reddening values are not always  
known, therefore increasing even more      
the uncertainty in assessing the temperature by scaling the synthetic flux
to the observed flux. As a consequence, for some targets we may present 
more than one model fit.  

The WD projected rotational velocity ($V_{rot} \sin(i)$) 
is determined by fitting the
WD model to the spectrum while paying careful attention to the 
line profiles.
We did not carry out separate fits to individual lines but
rather tried to fit the lines and continuum in the same fit while
paying careful attention to the absorption lines. 

For most spectra, when possible, we tried to fit 
a single WD model, a single disk model, and a composite WD+disk model,
assuming different reddening values. Systems that were observed mainly
in the low state can usually be modeled just with a WD spectrum,
while systems observed in a high state usually require the addition of 
a disk model.  
In systems for which the WD mass (and radius) is unknown, 
the distance is estimated using the maximum magnitude/orbital period relation 
\citep{war95,har04} 
or, when the period is unknown, we estimate the distance using the secondary
infrared emission technique  
of \citet{kni06,kni07}. 
  
For the single WD model, we vary the temperature while first keeping
the WD mass constant, starting at about $0.4M_{\odot}$.
Once the lowest $\chi^2$ has been found for a given mass, we 
vary the projected rotational velocity, and possibly also the abundances, 
to further lower the $\chi^2$ and obtain a best fit.  
Once the best fit has been found for that mass, 
we assume a slightly larger
mass and again vary the temperature until the lowest $\chi^2$ is found. 
We follow this procedure iteratively 
until we reach a mass of about $1.2M_{\odot}$.       
The next step is to find, from all these lowest $\chi^2$ models, the one which
agrees best with the distance estimate, or the one which has the
lowest $\chi^2$ of all (if the constraint on the distance cannot be
used). For the single disk model, we carry out a similar procedure by 
varying the mass accretion rate and inclination assuming discrete values
of the WD mass, and then choose the least $\chi^2$ model agreeing best
with the distance. We use a similar procedure 
for the WD+disk composite modeling to find the best fit model.  

In Table 6 we list the model fits we carried out for a number
of systems. These appear in the same manner as in the catalog. 
In the next section we concentrate on a few systems as examples
to illustrate the diversity of model fits.   

\section{Examples} 

We give here a few examples of systems exhibiting model fits
with increasing level of complexity. For the details of the model fits
of all the systems in Table 6, the reader is referred to the 
references in that Table.     

The simplest and easiest objects to model are those in which
the WD is the sole (or dominant) component at very low accretion
rates. In that case the spectrum resembles that of a field WD,
with little or no metal lines 
and with a low stellar rotational rate.  
This is the case for the pre-CV (a post-common envelope binary)
V471 Tau depicted in Figure 1. 
This eclipsing system has a magnetic WD accreting from the poles  
at a low accretion rate.  The continuum flux is characterized mainly 
by the absorption features of the Lyman series onto which are superposed
a few shallow metal absorption lines.  
The {\it FUSE} spectrum of V471 Tau was obtained as part of a 
Periodic Channel Coalignment program (rather than a ``usual'' 
scientific research GO program) and, as a consequence, the spectrum
had to be binned in time in a very specific manner to extract 
the data obtained only during the good exposure time (this is described
in detail in the next section). This {\it FUSE} spectrum has not been 
modeled previously. 
The results of the model fit to the {\it FUSE} spectrum are completely
consistent with the results obtained from fitting the STIS spectrum 
of the system at longer wavelengths \citep{sio12}, namely, the WD has a gravity $\log(g)=8.3$,
a temperature $T_{wd}=33,400$~K, sub-solar abundances (of the order
of 1\% solar and lower) and a broadening
of the absorption lines corresponding to 250~km~$s^{-1}$, about a factor 
of 4 larger
than expected, a possible indication that the broadening is due to 
a weak Zeeman effect.  

In all the online figures,  
the flux (vertical axis) is given in ergs$~$sec$^{-1}$cm$^{-2}$\AA$^{-1}$
and the wavelength (horizontal axis) is given in Angstr\"oms (\AA). For clarity 
all the spectra are shown on three panels: the upper panel from about 900~\AA\  
to 1000~\AA, the middle panel from about 1000~\AA\  to 1100~\AA, and the lower 
panel from about 1100~\AA\  to 1200~\AA. Air glow emission is annotated above 
each panel with a ``+'' sign inside a circle. Strong emission lines 
reaching the upper part of the frame in each panel are marked just 
above the panel.  The Hydrogen series is marked below the x-axis 
in the upper and middle panel and the hydrogen cut-off limit is 
clearly marked around 915~\AA\ with an increasing density of tick marks. 
The He\,{\sc i} (1168~\AA) line is marked below the lower panel. 
When a model is shown, the portions of the observed spectrum that 
have been masked before the spectral fit are in blue, while the 
rest of the observed spectrum is in red. The model fit is shown 
with a solid black line (on a white background) or a
solid white line (on a black background). 

As the mass accretion rate increases, the WD spectrum becomes more
affected by the metal lines. This is demonstrated in Figures 2 and 3. 
Figure 2 depicts the {\it FUSE} spectrum of the wind accreting WD P831-57. 
The model fit has a WD with $log(g)=7.85$, $T_{wd}=37,500$~K,   
with $1-2\%$ solar abundances and a slow rotational velocity. 
The details of this model fit are given in \citet{bar12}. 
In Figure 3 we show the spectrum of the nova-like MV Lyr
obtained in its low state, fitted with a solar abundance
WD model. The metal lines are much broader and deeper than
in Figures 1 \& 2. The temperature of the WD reaches 45,000~K. 
One can clearly see, from Figures 1-3, 
how the flux steadily increases in the shorter
wavelengths of {\it FUSE} as the WD temperature increases.   
Some cool 
WD models include the effect of quasi-molecular HI absorption features, 
in the catalog these are labeled ``Q-mol'' in the appropriate Figures.     

Not all the systems exhibit spectra that can be easily modeled
as emission lines (from the sources, or geo- helio- coronal in origin)
and ISM absorption lines can also impact the spectrum.
We show such a spectrum in Figure 4 for the symbiotic binary AE Ara.  
In the figures in the online catalog,  
neutral oxygen lines (when present)  
are shown with an arrow ``OI --$>$''; if many (ISM) OI lines are present, 
they are just annotated with vertical tick marks in the lower portion 
of each panel. ISM molecular hydrogen lines are either individually 
annotated vertically (e.g. L10R1, as in Figure 4) or just marked with an arrow 
``MH --$>$'' and vertical tick marks (usually in the upper part of each 
panel). Features resulting from fixed pattern noise introduced by 
{\it FUSE} detectors are indicated with ``FPN''.    

As a more complex example, we choose the {\it FUSE} spectrum of the 
dwarf nova (Z Cam) system EM Cyg, which is shown in Figure 5. 
The data were obtained in quiescence
and consists of four exposures taken at different binary orbital phase. 
We modeled the exposure obtained when absorption was minimal,
corresponding to the WD facing the observer. The S/N of a single
exposure is rather low and the spectrum is also affected by ISM
absorption. This system is dominated by emission from the WD but
a best fit is obtained with the inclusion of a low mass accretion 
disk model (see Figure 5) and ISM molecular absorption curtain. 
In the catalog, as in Figure 5,  
the disk model and the WD model  
are shown respectively with dashed- and dotted lines.

As a last example, we choose the UX UMa nova-like variable V3885 Sgr
observed in a high state, shown in Figure 6. This {\it FUSE} spectrum  
is different than the ones shown in the previous figures, 
and it is characteristic of systems accreting at a high accretion rate.  
The spectrum is characterized by a continuum flux extending to the shortest
wavelengths of {\it FUSE}, where the Lyman series absorption features are
not as obvious as in the previous spectra due mainly to Keplerian
rotational broadening. In addition, broad, deep, and blue-shifted
absorption lines of highly
ionized species (such as N\,{\sc iv}, S\,{\sc vi}, O\,{\sc vi}, .. ; see Figure 6),
are indicative of a hot corona forming above the disk or boundary layer(BL). 
The sharp absorption lines of H\,{\sc i} are of interstellar origin.  
Systems in high states are modeled mainly with an accretion disk with
a large $\dot{M}$ including the contribution from a hot WD, which
usually does not contribute much to the overall flux.
Here we improve the disk model, 
by including the contribution of the BL to
the theoretical spectra. We replace the inner disk rings with  hot 
rings matching the temperature of the BL $\sim 10^5$K. 
We built different BL models, consisting of 1, 2, 3 or more
rings assuming different ring temperatures ranging between 100,000K
and 200,000K. We generated models consisting of all the possible
combinations of BL + disk + WD, over a wide range of parameters 
keeping the assumption the BL luminosity should not exceed the disk
luminosity.  
We then apply our least $\chi^2$ fitting routine  
after we masked all the sharp ISM lines and the broad emission lines
which are all blue-shifted by $\sim$2\AA\  (an indication that
they form in a corona above the emitting component).  
The final (best) model is found based not only on the the least
$\chi^2$ value, but also by checking which (least $\chi^2$) model agrees
best with the distance, white dwarf mass, inclination, and mass 
accretion rate (if and when these are known).   
The best model to the {\it FUSE} spectrum of V3885 Sgr consists
of a 0.7$M_{\odot}$ WD with a temperature of 60,000K WD, 200km/s, 
a disk inclined at 60$^{\circ}$ with $\dot{M}=3 \times 10^{-9}M_{\odot}$/yr,
and a BL with a temperature of 175,000K (see Figure 6).   

We note that a hot WD and a high $\dot{M}$ disk also provide a satisfactory fit
\citep{lin09},
but in order to match the flux in the shorter wavelengths of {\it FUSE} one needs
to include the boundary layer.   

We are now implementing disk models with the inclusion 
of a realistic boundary layer to improve the
modeling of such systems in the high state.  
In the {\it FUSE} spectral
range, one has to include mainly the hottest components of the system,
namely the WD, the boundary (or spread) layer, and the accretion
disk.

\section{Acknowledgements} 
PG is pleased to thank the Henry Augustus Rowland Department 
of Physics and Astronomy at the Johns Hopkins University, Baltimore, MD, 
for hospitality. 
This work was supported by the National Aeronautics and Space Administration (NASA)
under grant number NNX08AJ39G issued through the Office of Astrophysics Data Analysis 
Program (ADP) to Villanova University. We have used some of the online data
from the AAVSO, and are  thankful to the AAVSO and its members worldwide for making
this data public and for their constant monitoring of cataclysmic variables.   
This work is based on observations made with the NASA-CNES-CSA Far Ultraviolet Spectroscopic Explorer. 
{\it FUSE} was operated for NASA by the Johns Hopkins University under NASA contract NAS5-32985.

\appendix

\section{Catalog Description} 

The {\it FUSE} catalog is maintained at the Mikulski Archive for Space Telescopes 
(MAST) as a High-Level Science Product (HLSP).   There is a search interface 
where users can specify query parameters to find objects of interest.   
From the search results, the user can bring up a webpage for each target 
that contains figures, downloadable data, system parameters and model parameters. 
The catalog data are stored in a relational database and the image 
components for the web pages are stored online.   
The  webpages for each object are built dynamically using PHP scripts 
using data retrieved from the database and the online images.  
This permits the catalog to easily accommodate growth, enhancements, and 
if necessary, corrections.  The FUSE mission search interface results 
also include links leading to the catalog thus providing general MAST 
users easy access.  The catalog is found at the URL:  http://archive.stsci.edu/prepds/cvaro/

\subsection{Definitions}

{\it{The target name}} is the name of the system as found 
in the MAST database.
For most of the systems it is the name as given in the \citet{rit03} 
Catalog of CVs (the most common objects). For
less common objects not included in the Ritter \& 
Kolb Catalog of CVs, it is the name as given in the
\citet{dow01} Catalog of CVs. When applicable we also 
give the additional name(s) of the object.   
For example, WD 0334-6400 is a DA+dMe binary in which the WD is undergoing
accretion of the companion star's wind. It is not a CV though it was
classified as one in the \citet{dow01} catalog.  
WD 0334-6400 is also known as Ret 1 or P831-57, 
and it is listed in our catalog under all three names.

{\it Type}: in the catalog we define four 
types of Catalysmic Variables as follows: 

DN: Dwarf Nova;  
\\
\indent 
NL: Nova-Like;  
\\
\indent 
Mg: Magnetic Systems;   
\\
\indent 
Msc: Miscellaneous variables.  

If only the type is entered (without any other selection), 
the search returns all the objects classified under this type.  

{\it Subtype}: the objects in the catalog are further divided into
subtypes as follows. 

DN subtype:
\\
\indent \indent  
UG - U Gem;  
\\ 
\indent \indent              
Z  - Z Cam; 
\\ 
\indent \indent              
SS - SS Cyg; 
\\ 
\indent \indent              
WZ - WZ Sge; 
\\ 
\indent \indent              
SU - SU UMa. 
  
NL subtype: 
\\
\indent \indent 
VY - VY Scl (anti-DN);  
\\
\indent \indent 
UX - UX UMa;  
\\
\indent \indent 
SW - SW Sex.

Mg subtype: 
\\ \indent \indent 
IP - Intermediate Polar (with a moderately 
strong magnetic field; disrupted inner disk);   
\\ \indent \indent 
P  - Polar (with a strong magnetic field; devoid of disk);   
\\ \indent \indent 
DQ - DQ Her; 
\\ \indent \indent 
AM - AM Her.

Msc subtype: 
\\ \indent \indent 
N  -  Nova (all types of novae);  
\\ \indent \indent 
CV - all other type/subtype of Cataclysmic Variables;  
\\ \indent \indent 
R - related to CVs: Symbiotics;    
\\ \indent \indent 
PCV - Pre Cataclysmic Variable.

If only the subtype is entered (without any other selection), 
the search returns all the objects classified under this subtype.  

{\it P} is the orbital period of the binary (if known) given in hr.  

{\it i} is the inclination of the system (if known) and is given in
degrees. It is also the inclination used in the modeling
of the synthetic disk spectrum to fit the observed spectrum
(it ranges between 0 and 90).  

{\it E(B-V)} is the reddening (extinction) toward the
object. The spectrum was dereddened assuming this
value prior to the spectral fit. 
Unless otherwise specified, the reddening value is taken for
each system from the works of \citet{ver87,lad91,bru94}.

$M_{wd}$ is the mass of the WD of the system given in units of
solar mass. This mass is usually taken from the literature
and the entry is left blank if it is unknown. In the modeling
we use the $log$ of the surface gravity ($log(g)$) rather than the mass 
of the WD;  
it corresponds to the effective surface gravity (acceleration in $log$ 
and cgs units) of the WD used to generate the stellar
atmosphere or accretion disk model to fit the observed spectrum  
(it ranges between 7.0 and 9.5).  
 
$T_{wd}$ is the effective temperature of the WD, 
given in Kelvin, that was used to generate the stellar atmosphere
model to fit to the observed spectrum  
(usually between 10,000~K and 100,000~K).  

$V_{rot} \sin{i}$ (online as {\it Rotv}) 
is the projected rotational velocity of the WD
(in km~$s^{-1}$) used to generate the line broadening of the
stellar atmosphere model in the fit to the observed spectrum, 
it usually ranges between 10s~km~$s^{-1}$ to a few 100~km~$s^{-1}$.  

{\it d} is the distance to the object given in parsecs and
it is the distance used in the spectral
modeling to fit the observed spectrum. 
If the distance is not known from the literature,  
then the work of \citet{kni06,kni07} is used to gain information on the
secondaries and to estimate a lower limit on the distance.

{\it C} is the abundance of carbon in solar units 
as derived from spectral modeling.
 
{\it Si} is the abundance of silicon in solar units 
as derived from spectral modeling.

{\it S} is the abundance of sulfur in solar units 
as derived from spectral modeling.

{\it N} is the abundance of nitrogen in solar units 
as derived from spectral modeling.

{\it Abundances} are the abundances of all the species: either 
solar, or non-solar or as specified (e.g. N/C$>$10).   

The mass accretion rate is given in the modeling in $log$ units of solar 
masses per year. This is the mass accretion rate
that was used in the spectral disk modeling to fit the
observed spectrum, and  
it usually ranges anywhere between -12.0 and -8.0.    

\subsection{Search Page}
 
One can either search for an object using the usual search keys (such as 
Target Name, RA \& Dec, ..), or by Type and Subtype (returns all objects listed
in that category). One can also use the User-specified fields 1 \& 2 as follows. 
Choose the User-specified field 1, e.g. $T_{wd}$ for the WD temperature, and enter
in the Field Description the lower limit for Twd in Kelvin, e.g. '$>$19999' (without
the quotation mark and enter the thousands without a coma); 
do the same with the User-specified field 2 for the upper limit for Twd,  
e.g. '$<$30001'. Hit search and all the systems with a WD temperature between
20,000~K and 30,000~K are returned. The temperature returned here is the one 
obtained from the modeling or known from the literature.  
If the search button is hit without filling in any key, the entire list of
objects is returned. Note: the other buttons (e.g. maximum records,
records per page,...) are the ones usually found in any search on MAST.  
The object(s) from the search is(are) returned in a table (``Search Results Page''). 

\subsection{Search Results Page} 

The Search Results Page returns the objects (Targets) as a table. 
The table contains from left to right: 
\\
- Mark: this button marks the dataset for which all the available files  
(spectrum, model and figure) can be downloaded by pressing the 
``Download marked data'' button (use this button only if you want 
more than just the {\it FUSE} spectrum). 
\\
- Target Name. By clicking on the name 
of the target the ``Object Page'' is returned. This is the most important 
link for each object!   
\\
- RA and Dec (in Julian 2000). 
\\
- Type.  
\\
- Subtype. 
\\
- The Period. 
\\
- The inclination.  
\\
- The reddening. 
\\
- The WD mass.
\\
- The WD effective surface temperature.
\\
- The projected rotational velocity.
\\
- The abundances of Carbon (C), Silicon (Si), Sulfur (S), Nitrogen (N) and
Abundances in general. 
\\
- The distance.
\\
- Data ID in last column is just a key for the software (no use for the user).  

In Table 1 we show a table similar to the search results page, in which 
we have omitted the WD temperature, 
projected rotational velocity and abundances.  


\subsection{Object Page}  

The object page displays all the information, data 
and graphics for an individual object (system). 
This is that page you will  
want to access when you are 
interested in a particular object.  
The Object Page is obtained (or 'returned', or 'accessed') by 
clicking directly on the name of an object in the Search Results Page
table.  The object page is divided into several panels. 

{\bf The Description Panel}. 
The upper panel is the ``description'' panel, with the name 
of the object in bold. On the left, it contains the basic parameters for that 
object: RA, Dec, Type, Subtype, $P$, $i$, E(B-V). 
On the right, this upper panel displays a basic 
description of the
{\it FUSE} spectrum/spectra, the modeling/s and the figure/s displayed below. 
In the bottom of the upper panel there is a direct link to a list of references 
(``bibliography'') for that object. That list is not complete as it only
contains the most important references and the references relevant to the 
{\it FUSE} spectra and its modeling for that object.  
Do not hesitate to contact us to suggest additional references.

{\bf The Figure(s)}.  
For most objects there is only one additional panel located under the 
upper (description) panel. In that panel, on the right, there is a
figure (in GIF format) displaying the {\it FUSE} spectrum (in red) of the object
together with a synthetic (model) spectrum (in white or black). Line 
identification is  included 
in the figure (for a description of the lines see ``The {\it FUSE} Spectral Lines''
below). By clicking on the figure one can view the figure by itself, and by
clicking on it again a full resolution of the figure is returned (enlarged). 
In most cases the features in the spectrum that are not associated with
the source (e.g. ISM lines, geocoronal emission, etc..) are marked in
blue and masked before the modeling is carried out. 
Under the figure is listed the reference in which the original data
analysis and modeling were carried out. 

{\bf Observation Log Display}. 
On the left to the
figure, under the title {\it FUSE} Data, 
information on the {\it FUSE} spectrum is given (DataID, Observation
Date and Time, good exposure time, state in which the system was observed: 
HIGH/OUTBURST, LOW/QUIESCENCE, or INTERMEDIATE) 
together with a direct
link (``Download {\it FUSE} data'') to the {\it FUSE} spectrum in ascii format
(a table with wavelength, flux and error; the table contains a header
with basic info; the units are as shown in the figure).   

{\bf Model Fit Display}. 
Below this, under the title Synthetic Spectrum,  
the parameters that were used to model the spectrum are given, e.g.
whether it is a simple disk model (DISK), or a WD, or a combination
of both (WD+Disk), together with the quantities used (inclination, 
reddening E(B-V), gravity $log(g)$, temperature (K), projected rotational velocity
(km~$s^{-1}$) of the WD, C, Si, S and N abundances in solar abundances, distance
in pc, mass accretion rate ($log(\dot{M}$) in $M_{\odot}$/yr). Below that is a
direct link to the model (``Download Synthetic Spectrum'') which is given
in ascii format as a table with wavelength and flux. The ascii table
of the model contains a short header with basic information and the units are
as shown in the figure. 

Some systems were not modeled and do not have  a synthetic spectrum and, 
therefore, that part under ``Synthetic Spectrum'' has no data. 

Some systems have more than one {\it FUSE} spectrum and this is reflected by 
having more than one data block under the title {\it FUSE} data and/or 
(possibly) more than one figure.  
 
Some systems have more than one model and this is reflected by
having more than one data block under the title Synthetic Spectrum
and/or more than one figure. 

{\bf Notes On Figures}. 
The bottom panel in the Object Page contains general notes on the figures. 
These notes can be found here in section 5.

{\bf{References for Individual Objects} } 

We list references for each object in Table 7. 
In the online catalog, 
references for each object are given in the upper part of the
object page under the link ``Bibliography''. This does not cover all the
existing references for the object as some objects have been extensively
studied. If you find that an important reference is missing, please
contact us ( patrick.godon@villanova.edu ) and we will incorporate any
additional relevant reference you provide. Do not hesitate to contact us for
comments, error and suggestions. 

For the IP systems, instead of providing a list of references, we provide
for each IP a direct link to the system in the online catalog of IPs of
Koji Mukai\footnote{http://asd.gsfc.nasa.gov/Koji.Mukai/iphome/iphome.html 
(this is a case sensitive URL). }.    

For each system, a link is also given under each figure 
to the work in which the data analysis was
originally carried out. That reference usually contains the full details
of the data processing and modeling and is listed here for each system
separately in Appendix B.

\section{Individual Objects} 

In this section, we provide notes on individual objects that are included
in the initial release of the online catalog. The search function in the 
online catalog returns the objects listed in alphabetical order, and
for this reason we list here the objects in alphabetical order.

\paragraph{AE Aquarii} 
The spectrum of the Intermediate Polar AE Aqr shows 
broad and strong emission lines of N\,{\sc iii} + He\,{\sc ii} ($\sim$990), 
O\,{\sc vi} (doublet), and He\,{\sc ii} (1085). 
We note, however, the absence of the usually 
strong C\,{\sc iii} (977 \& 1175) emission lines; instead, we tentatively
identify these lines with the weak spectral features there. 
Weak broad emission lines 
of S\,{\sc iv} (1063 \& 1073) and N\,{\sc iv} (923) are detected. 
The S\,{\sc vi} (933 \& 944) emission lines which are sometime seen in
systems exhibiting a variety of rich emission lines (e.g. such as e.g. DW UMa 
\citep{hoa03}) are not detected at all but they are marked on the figure. 
The {\it FUSE} exposures cover the entire orbital phase 
and the spectrum was obtained by combining 
the exposures obtained when the emission from the poles is minimal. 
However, the contribution from the white dwarf produces only 
a very weak continuum,
while the matter shocked at the poles emits the broad 
and strong emission lines.     
This object exhibits the N/C anomaly indicative of CNO processing,
as first noted in studies at longer UV wavelengths \citep{gan03}. 

\paragraph{AE Arae} 
The {\it FUSE} spectrum of the Symbiotic AE Ara (Figure 4) 
shows many ISM absorption lines,  
in particular the molecular hydrogen lines. The very sharp emission lines of 
the Hydrogen Lyman series and O\,{\sc i} are most probably geo- and 
heliocoronal in origin. 
The sharp C\,{\sc iii} (1175), S\,{\sc vi}, S\,{\sc iv} lines are from the source, though some of 
the sharp lines might be contaminated with sunlight reflected in the SiC channels (e.g. C\,{\sc iii} 977).   
We note that the S\,{\sc vi} (933, 944) and the oxygen doublet appear to have two components:
the first is in emission at about the rest wavelength, and the second is in absorption with 
a blue shift of at least 2\AA\ . 

\paragraph{AG Draconis} 
The spectrum of the Symbiotic AG Dra is characterized by sharp emission lines 
from highly ionized elements from the nebula being ionized by the hot component(s) 
of the system. 
These lines include Ne\,{\sc v}, Ne\,{\sc vi}, Ne\,{\sc vii} 
lines as well as lines from the He\,{\sc ii} Balmer series. 
S\,{\sc iv}, S\,{\sc vi} sharp emission 
lines are also seen as well as lines of Fe\,{\sc ii} (1141.17) and 
Fe\,{\sc iii} (1142.43) fluoresced 
by the very broad O\,{\sc vi} (sharp O\,{\sc vi} emission
lines are also present). The ISM absorption lines include 
Si\,{\sc ii}, Fe\,{\sc ii}, N\,{\sc i} and Ar\,{\sc i} lines. 
The rest of the continuum is rather {\it flat} indicative of a very
hot component.  
 
\paragraph{AM Cassiopeiae} 
AM Cas was observed with {\it FUSE} in an intermediate state of brightness.
The longer wavelengths of the spectrum are strongly affected by the
worm.    
The spectrum was modeled with WD+disk model assuming both E(B-V)=0 and E(B-V)=0.2. 
The distance to the system is possibly around 350~pc. 
The mass of the WD is unknown and so is the system inclination. 
We assessed $M_{wd}$ and $i$ here from the modeling. 
Note the difference between model 1 and model 2 in Table 6.  
Details on the processing of the observed spectrum and
the spectral analysis are given in \citet{god09a}. 
The two models shown in the Figures (online) do fit the longer
wavelengths of the {\it FUSE} spectrum but in the shorter
wavelengths the observed flux is larger than in the models. 
The excess of flux could be due, for example, to broad emission
from N\,{\sc iv} (923) and S\,{\sc vi} (933, 944) but as shown in the
figures the observed features are not at the expected wavelengths
and do not have the proper line profile, in addition excess of flux
also occurs around 950 to 960 \AA\ , and the line notation is
therefore only suggestive. Lines such as Si\,{\sc iii} (1110) and 
Si\,{\sc iv} (1120-1130) make large shallow depression which fit
the disk-dominated model, indicative that Keplerian broadening is
responsible for their appearence. The ISM absorption lines (molecular hydrogen) are colored
blue in the Figures so as to not be confused with the other lines. 

\paragraph{AM Herculis} 
The {\it FUSE} spectrum of the Polar AM Her, obtained in its low state,
consists of 29 individual exposures, 
giving a complete coverage of the orbital phase of the binary (from 0 to 1). 
There are some sharp but weak ISM molecular hydrogen absorption lines as well 
as some metals: Fe\,{\sc ii}, Ar\,{\sc i}, N\,{\sc ii}, C\,{\sc ii} and P\,{\sc ii}. 
The O\,{\sc vi} lines are possibly due to airglow.  We mark the position of the 
C\,{\sc iii} (1175) line to emphasize its absence. The WD is clearly visible 
and is the main source of FUV light. 
The combined spectrum is available for download as well as  
the individual exposures (1 through 29).

\paragraph{AQ Mensae}
Due to their contamination with air glow and very low S/N, the LiF 1 and SiC 
1 \& 2 channels have been discarded. 
The system is always at about the same magnitude (14-15). 
Details on the processing of the observed spectrum are given in \citet{god09a}.  
The S/N is so low that one cannot identify lines. However, some of the usually
most prominent ISM molecular hydrogen absorption lines coincide with some
low flux regions in the spectrum, indicating that the features might be real 
(e.g. L5R0, L5R1, L5R2, L4R0, L4R1, L4P1, L4R2, .. through L0R0, L0R1, L0P1). There is
also a spectral feature that could be associated with the C\,{\sc iii} (1175)
line, though the overal flux level in that region is not reliable. Lines have been
marked in the online Figures but are not identified with reliability, they are only
suggestive. 

\paragraph{BB Doradus} 
BB Dor was observed with {\it FUSE} in a high state. We fit here four different models 
as shown in Table 6: a WD, a disk, a WD+Disk at low inclination, and a WD+Disk 
at high inclination. 
The high inclination disk contributes much less flux than the WD, while 
the low inclination disk contributes much more flux than the WD.  
The sharp emission lines, all colored in blue, are most likely due to the reflection
of day and/or sun light in the telescope and therefore marked with a C in Table 4. 
There are lines that appear clearly in the model but cannot be identified with
reliability in the observed spectrum; these lines are marked on the figure but 
appear with a D in Table 4 (e.g. S\,{\sc iv} 1006 \& 1100). The S\,{\sc iv} 
(1063 \& 1073) and Si\,{\sc iii} (1110) lines are much deeper in the observed
spectrum than in the model, possibly a sign that there is additional absorption from
material above the disk. The oxygen doublet is also in absorption and is not modeled. 
All the details of the spectral modeling, data analysis and processing
of the data are given in \citet{god08a}.  

\paragraph{BV Centauri} 
The {\it FUSE} Spectrum of the U Gem dwarf nova BV Centauri was 
taken in quiescence. The spectrum is very noisy especially at the 
shorter wavelengths. The increase of flux at 
at 920~\AA\  is probably due to a background subtraction problem. 
We identify ISM Hydrogen molecular absorption features as well as some
ISM iron lines. 
The sharp emission lines are due to terrestrial airglow. 
The broad O\,{\sc vi} doublet and C\,{\sc iii} (977~\AA ) are from the source. 
The S\,{\sc iv} (1073), Si\,{\sc iii} (1113),  Si\,{\sc iv} (1123 \& 1128) 
and C\,{\sc iii} (1175) absorption spectra features are all identified. 
All the other absorption features marked on the figure are suggestive and are
not detected. 
The modeling takes into account the larger mass recently found of $1.24 M_{\odot}$ 
by \citet{wat07a}, this does not change the temperature of the WD appreciably, 
but it reduces the distance to about 200~pc. The modeling includes the
effect of the ISM and the theoretical spectrum that is compared to the observed
one consists in the spectral regions colored in red (in the second online figure). 
Details on the processing and spectral modeling of the {\it FUSE}
spectrum are given in \citet{sio07}.  

\paragraph{CH Ursa Majoris} 
The {\it FUSE} Spectrum of the U Gem Dwarf Nova CH Ursa Majoris was taken in quiescence. 
The spectrum is very noisy especially in the shorter wavelengths. 
Due to the hydrogen cut-off one expects the flux to drop significantly 
for wavelengths $ < 915$~\AA , instead the flux keeps on increasing, 
a sure sign of a noisy detector edge as well as a possible background 
subtraction problem. We identify ISM H molecular absorption features 
as well as ISM metal lines, mainly of Fe\,{\sc ii}. 
The spectrum is also affected by terrestrial airglow (sharp emission lines) 
and the S\,{\sc vi} (944) feature, if real, is also probably due to day light
reflected in the telescope. 
There are some broad emission features that are associated with the source: 
we identify the O\,{\sc vi} doublet and C\,{\sc iii} (977 \& 1175~\AA ). 
All the other lines marked in the figure are not detected. 
In the modeling we add an ISM model to mimic the absorption due to the 
Hydrogen molecular absorption features. 
Details on the processing and spectral modeling of the {\it FUSE}
spectrum are given in \citet{sio07}.  

\paragraph{DQ Herculis} 
The {\it FUSE} spectrum of the IP DQ Her has strong and broad 
emission lines. These include 
N\,{\sc iv}, S\,{\sc vi}, C\,{\sc iii}, He\,{\sc ii}, and O\,{\sc vi} lines. 
The continuum of the flux is very low and shows some weaker 
emission lines from P\,{\sc iv}, P\,{\sc v}, S\,{\sc iv} and Si\,{\sc iii}.  
The emission lines originate near the poles where matter accretes through
magnetic lines.  

\paragraph{DT Apodis} 
DT Aps was observed twice with {\it FUSE}, but only the second {\it FUSE} spectrum 
shows a continuum. This low S/N spectrum is heavily affected by air glow, especially 
at the shorter wavelengths. The deep absorption feature (1110-1130~\AA ) is probably 
due to an incorrect background subtraction (as is probably the C\,{\sc iii} feature). 
One cannot identify with reliability any spectral line from the source. 
The system parameters are unknown and, therefore,  one cannot model the spectrum 
as there is a degeneracy in the modeling.   
Details on the processing of the observed spectrum 
are given in \citet{god09a}.  

\paragraph{EK Trianguli Australis} 
The {\it FUSE} spectrum of the DN EK TrA in quiescence is modeled with a WD. 
The effect of the Hydrogen quasimolecular opacity has to be included to agree
with the flux around 1060-1080~\AA . The theoretical spectrum without the 
inclusion of the H quasimolecular absorption is shown with a dashed line
in that region (1060-108 \AA ). All the sharp emission lines (including the nitrogen
lines) are due to airglow. The broad lines of C\,{\sc iii} (977 \& 1175) and 
O\,{\sc vi} are from the source. In the longer wavelengths of {\it FUSE}
the theoretical and observed spectra match each other in some regions
but not everywhere. For example the S\,{\sc iv} has an observed flux
lower than modeled at 1123\AA\ , while at the bottom of the broad
absorption feature at 1128\AA\ there is a small sharp emission feature
most probably due to the low S/N. These lines are identified by their
location and the by the overall shape of the continuum in their immediate
neighborhood. 
The spectrum is a combination of two {\it FUSE} spectra, 
and has been dereddened assuming the published value E(B-V)=0.03. 
Details of the modeling are given in \citet{god08b}. 

\paragraph{EM Cygni} 
The {\it FUSE} data of EM Cyg were obtained during a relatively low state of activity. 
The spectrum (see Figure 5) 
consists of a {\it FUSE} exposure taken at orbital phase 0.15 when the flux
was maximum. The other {\it FUSE} exposures have a lower flux.
The {\it FUSE} spectrum is strongly affected by ISM H molecular absorption
which is included in the modeling. 
The N\,{\sc iv} (923) and S\,{\sc vi} (933) lines are not identified due to the 
low S/N, and S\,{\sc vi} (944) could possibly be in absorption. 
The C\,{\sc iii} (977 \& 1175) and O\,{\sc vi} lines are seen each in absorption on
top of a broader emission feature. The only lines that we identified in absorption
(and are also deeper than in the model) 
are S\,{\sc iv} (1073), Si\,{\sc iii} ($\sim$1110) and Si\,{\sc iv} (1123 \& 1128). 
The other lines that are usually observed are marked on the figure but are not identified. 
The model spectrum consists of a WD+Disk. 
The WD component contributes 93\% of the flux, the disk contributes the remaining 
7\%. Some absorption lines cannot be accounted for and are possibly due to veiling 
of the WD from the L1 stream material flowing over the disk to smaller radius at different 
orbital phases. Maximum veiling is around orbital phase 0.6-0.9.   
All the details of the processing of the {\it FUSE} spectra and their analysis is
given in \citet{god09b}. 

\paragraph{ES Draconis} 
ES Dra was observed in an intermediate state and it is not clear whether the 
main component is a WD or a disk, as the inclination and WD mass of the system are not known.   
Because of that, the {\it FUSE} spectrum of ES Dra is modeled first with a single WD, and then with a 
DISK alone. We do not consider any WD+disk model, as there is a degeneracy of the solution 
(there are many WD+disk models that can fit this spectrum covering a large area in the parameter space).  
Details on the processing of the observed spectrum and
the spectral analysis are given in \citet{god09a}.  
The spectrum exhibits some sharp emission lines due to air glow as well as 
ISM absorption lines (mainly molecular H, but also some metals). 
There could be some broad emission near 977\AA  (C\,{\sc iii} ?) as well 
as near 990\AA . The C\,{\sc iii} (1175) feature seem to have blue shifted 
absorption component with a red shifted emission component. 
We identify with some reliability absorption lines/features from 
C\,{\sc ii} (1010),Si\,{\sc iv} (1066), 
Si\,{\sc iii} ($\sim$1110 \& $\sim$1143), Si\,{\sc iv} (1123-1128), as these lines are located
at the expected wavelengths and agree grossly with the model.
The S\,{\sc iv } (1063, 1073) lines are contaminated with ISM absorption. Some of 
the lines that are marked on the figure are present in the theoretical spectrum
but are not detected in the observed spectrum (e.g. S\,{\sc iv} $\sim$1006 
\& 1100). 

\paragraph{EX Hydrae} 
The IP EX Hya was observed with {\it FUSE} during
both a lower and higher state, as it changed with the orbital phase. 
The spectrum here was obtained in the low state in which the WD               
is possibly revealed and contributes a significant fraction of the FUV flux. 
The spectrum has relativelty good S/N. There is a sharp hydrogen Lyman$\beta$
emission line due to air glow. C\,{\sc iii} (977 \& 1175) and the O\,{\sc vi}
doublet are characterized by a shallow absorption on top of a braoder emission.
The oxygen doublet also exhibits a narrow emission slightly red shifted. 
The atomic hydrogen and oxygen narrow absorption lines are  from
the ISM, and so are the C\,{\sc ii} (1036 \& 1037) and nitrogen lines (though
some of the nitrogen lines could be contaminated from the terrestrial atmosphere).  
The system exhibits narrow absorption lines (broader than the ISM lines) 
of highly ionized species such as 
S\,{\sc vi} (933, 944), P\,{\sc iv}, P\,{\sc v}, O\,{\sc vi}, S\,{\sc iv}, Si\,{\sc iv} as well 
as lines of Si\,{\sc ii}, Si\,{\sc iii}, S\,{\sc iii}, S\,{\sc ii}, C\,{\sc ii}, C\,{\sc iii} and 
He\,{\sc ii}. 
This indicates that there are several absorbing layers with different temperatures. 

\paragraph{EY Cygni} 
This low S/N {\it FUSE} spectrum of EY Cyg was obtained in quiescence and is 
modeled with a WD spectrum. 
The increase of flux in the short wavelengths ($<$950~\AA) is an artifact 
in one of the {\it FUSE} channels due to inaccurate background subtraction, 
the N\,{\sc iv} (923) lines have been marked but are not detected and do
not match any observed feature. 
All the sharp emission lines of H, O, He) are due to geo- and helio-coronal
emission (light reflected in the telescope). There could be some 
broad emission from O\,{\sc vi} (1032) and, though less likely, from O\,{\sc vi} (1038). 
We have marked some of the ISM absorption lines where they 
match some absorption features in the observed spectrum such as some hydrogen
molecular lines and C\,{\sc ii} (1036). With some low (but non-zero) reliability
we identify some absorption lines matching the theoretical spectrum as follows:
S\,{\sc iv} (1073), Si\,{\sc iv} (1128) and Si\,{\sc iii} ($\sim$1142-1145). 
All the other lines that are marked on the figure are not identified and are there
mostly for comparison and identification with the theoretical spectrum. 
The {\it FUSE} spectrum of this system was processed and modeled in 
\citet{god08b}. This object exhibits the N/C composition anomaly indicative
of CNO processing.  

\paragraph{FO Persei} 
FO Per was observed
with {\it FUSE} in a high state and was modeled with an accretion disk 
and a heated WD, assuming both E(B-V)=0.0 and E(B-V)=0.3. 
Details on the processing of the observed spectrum and
the spectral analysis are given in \citet{god09a}.  This {\it FUSE} spectrum
is rather noisy and has been binned to a lower resolution for slightly
increase the S/N. All the sharp emission lines are most likely day light
contamination (all the H\,{\sc i} lines, the S\,{\sc vi}, O\,{\sc vi}, 
C\,{\sc iii} and He\,{\sc i} lines). We identify only the Si\,{\sc iv} (1066), 
Si\,{\sc iv} (1123 \& 1128) and C\,{\sc iii} lines with the source. All the other lines
that are marked are either contaminated with ISM absorption (mostly molecular
H), terrestrial in origin (e.g. N\,{\sc i}), or are of the order of the noise. 
All the features that are not from the source have been marked in blue
in the figure and were masked before the modeling (spectral fit). 

\paragraph{HP Normae} 
HP Nor was observed in a relatively high state and was modeled with an 
accretion disk. The {\it FUSE} spectrum was dereddened assuming E(B-V)=0.2. 
Details on the processing of the observed spectrum and
the spectral analysis are given in \citet{god09a}.  
The {\it FUSE} spectrum has a very low S/N and is contaminated with sharp
emission lines (geo- and helio-coronal in origin). The ISM molecular hydrogen
absorption features are slicing the spectrum at regular intervals. The only
feature that can can be associated with the source and that somehow
matches the model (withint the S/N range) is the C\,{\sc iii} (1175) multiplet. 
The ISM absorption features and the day light contamination have been marked
in blue in the figure and have been masked before the fitting. 

\paragraph{IX Velorum} 
In this high S/N {\it FUSE} spectrum of IX Vel,  
the sharp absorption lines are all observed at their theoretical rest wavelengths, 
while the broad absorption lines are blue shifted by about 2~\AA\  
and exhibit a sharp steep red wing and a much broader blue wing. 
All the broad absorption lines are associated with the source and have been
marked on the figure at the rest wavelength such that the blue shift is evident. 
All the sharp absorption lines are from the ISM, except possibly for the nitrogen
lines (terrestrial). 
Processing of the {\it FUSE} spectrum is given in \citet{lin07} together
with a spectral modeling of the {\it FUSE} spectrum combined with a STIS and 
IUE spectra.  

\paragraph{MU Camelopardalis} 
The {\it FUSE} spectrum of the IP MU Camelopardalis is of a very low quality. 
All the sharp emission lines are airglow contamination, possibly including 
S\,{\sc vi}, C\,{\sc iii} and O\,{\sc vi} from sunlight reflected on the SiC channels. 
Though the S/N is low, we clearly identify ISM molecular hydrogen, as 
well as absorption lines from C\,{\sc ii}. 
No S\,{\sc iv} lines are detected. Except for the broad O\,{\sc vi} 
doublet emission lines, no other spectral feature can be associated with the source. 
The continuum is rather flat and could indicate little contribution from the accreting WD.  

\paragraph{MV Lyrae} 
This {\it FUSE} spectrum of the Nova-like (VY) MV Lyr was taken during a low state. 
The spectrum (see Figure 3) 
reveals the accreting WD, which is heated due to 
continuous accretion as typical for NL VY systems. 
There is no discernable extinction toward MV Lyr. 
The spectrum consists of four exposures ({\it FUSE}  orbits). 
The WD model (solid white line) has a mass of $0.73~ M_{\odot}$,
a temperature of 45,000~K, a rotational velocity of 200~km~$s^{-1}$ and solar composition. 
To fit the flux below 930~\AA\ one has to decrease the temperature to 44,000~K. 
The slight decrease of flux (between 915 and 930~\AA ) is possibly due to the sharp 
H\,{\sc i}  absorption lines. In order to fit the right wing of the Ly$\beta$ 
profile one has to increase the temperature to 47,000~K. The right wing might 
be affected by some O\,{\sc vi} emission. 
All the sharp emission lines (H\,{\sc i}, C\,{\sc iii}, and more) are most
probably from air glow. We identify the following absorption lines which fit 
the WD photosphere model: S\,{\sc iv} (1006), C\,{\sc ii} (1010), S\,{\sc iv} (1063, 1073), 
Si\,{\sc iv} (1066), He\,{\sc ii} (992, 1085), Si\,{\sc iii} (1110), P\,{\sc v} (1118), 
Si\,{\sc iv} (1123, 1128) and C\,{\sc iii} (1175). All the other lines are either not
identified or they are sharp ISM absorption lines (e.g. Si\,{\sc ii}, C\,{\sc ii}, Ar\,{\sc i},
Fe\,{\sc ii}). 
For solar composition, the WD has 
to have a rotation rate of at least 200~km~$s^{-1}$ and possibly as fast as 250~km~$s^{-1}$ in order to match 
the profile of the absorption lines. If one decreases the abundances, the rotational velocity 
needed to match the absorption lines decreases as well. For Z=0.2 the (projected) 
rotational velocity needed to match the lines is 150~km~$s^{-1}$. For an inclination 
of 12$^{\circ}$, a projected rotational velocity of 200~km~$s^{-1}$ corresponds to 
about 1/3 of the Keplerian velocity at one stellar radius. For a velocity 
of 150~km~$s^{-1}$, it corresponds to about 1/4-1/5 of the Keplerian velocity. 
Note that if one adopts $i=7^{\circ}$ as in \citet{lin05} then 250~km~$s^{-1}$ 
corresponds to 2/3 of the Keplerian velocity. It is more likely that the 
inclination is 12$^{\circ}$ and the projected rotational velocity 150~km~$s^{-1}$, 
and that the WD rotates at a fraction of about 1/4 to 1/5 of the breakup velocity. 

\paragraph{NSV10934} 
NSV 10934 was observed with {\it FUSE} twice consecutively, and we combined the 
two individual spectra.  At the time of the observations NSV 10934 was in 
a low state about a month after an outburst. The low S/N (especially in the 
lower wavelengths) is due to the fact that the data from the 2bSiC, 2bLiF 
and 1aSiC channels were unusable and a large portion of the 1bLiF channel 
was lost to the infamous worm. Because of this there is a gap around 1085~\AA\  
and the spectrum is not modeled.  
Details on the processing of the observed spectrum can be found
in \citet{god09a}.  The only spectral feature that we identify and which
belong to the source are the braod emission from C\,{\sc iii} (977 \& 1175) 
and the O\,{\sc vi} doublet. The other spectral features that are identified 
are the ISM molecular hydrogen absorption lines. All the other other spectral
lines marked on the figure are only suggestive. 

\paragraph{P831 57}    
The {\it FUSE} spectrum of the DA WD + dMe binary P831 57 was taken as part of a 
NL {\it FUSE} survey as the system was classified as a NL. From the present analysis 
it was found that the system does not have a disk and the WD is accreting 
from the wind of the secondary. The system also appears in the Downes 
catalog as NL called Ret 1, and it is also known under 
the name WD 0334-6400. 
The spectrum together with the model is shown in Figure 2 and details
on the analysis can be found in \citet{bar12}. 
On the figure the following lines are marked but are NOT identified: 
N\,{\sc iv} (923), S\,{\sc vi} (933, 944), O\,{\sc vi} doublet. All the sharp
emissin lines are either helio coronal or air glow. The spectral feature around
1152\AA\  is a well known fixed pattern noise (FPN) in {\it FUSE}, a smaller FPN feature
is also detected around 1072\AA\ . Since the WD has an extremely low rotation, its 
absorption lines are very sharp, similar to ISM lines. However, only the WD lines 
match the theoretical spectrum: C\,{\sc iii} (977, 1175), N\,{\sc iii} + He\,{\sc ii} (990),
Si\,{\sc iii} ($\sim$995), S\,{\sc iv} (1006), C\,{\sc ii} (1010), C\,{\sc ii} (1036), 
S\,{\sc iv} (1063, 1073), Si\,{\sc iv} (1066), N\,{\sc ii} + He\,{\sc ii} ($\sim$1085), 
Si\,{\sc iii} (1008, 1010, 1013), Si\,{\sc iv} (1023, 1028) and N\,{\sc i} (1134). 
Some of these lines are deeper in the observed spectrum than in the model, 
possibly an indication that they are superposed to ISM lines (e.g. C\,{\sc ii} 1036,
N\,{\sc i} 1134). All the other sharp lines are from the ISM (N\,{\sc i} $\sim$954,
Ar\,{\sc i}, etc..). 

\paragraph{RU Pegasi} 
This {\it FUSE} spectrum is rather noisy and is heavily affected by ISM absorption lines
from atomic and molecular hydrogen, as well as from metals (Ar\,{\sc i}, Fe\,{\sc ii}, 
S\,{\sc i}, N\,{\sc i}). The C\,{\sc iii} (977 \& 1175) and oxygen doublet lines are seen
in broad emission with a narrower absorption feature superposed to it. 
We identify absorption lines from N\,{\sc iv} (923), S\,{\sc vi} (933, 944),
oxygen doublet, and C\,{\sc iii} (1175) which are all slightly red shifted (by about
1\AA\ ), an indication that they possibly do not belong to the WD photosphere per se. 
We do, however, identify some absorption lines from the WD that are not shifted: 
Si\,{\sc iv} (1066), O\,{\sc iv} (1068) and S\,{\sc iv} (1073). The O\,{\sc iv} 
line do appear in very hot WDs and is a sign that the WD in RU Peg is indeed relatively
hot and is usually not observed in accreting WDs in CVs. 
The {\it FUSE} spectrum of the DN RU Peg in quiescence is modeled with a WD spectrum. 
The very high temperature of the model is needed to fit the flux level of a massive 
(and therefore compact) WD (i.e. to fit the distance) and to account and fit properly 
the S\,{\sc iv} (1073), Si\,{\sc iv} (1066), and O\,{\sc vi} (1068) absorption lines.    
The {\it FUSE} spectrum of this system was processed and modeled in 
\citet{god08b}. 

\paragraph{RW Sextantis} 
The spectrum of RW Sex consists of 25 {\it FUSE} exposures (orbits) combined together. 
From exposure to exposure the continuum remains fairly constant and only the 
bottom of the broad lines shows some variation. The spectrum shows broad absorption 
lines from high order ionization species (all blueshifted here by about 3~\AA): 
N\,{\sc iv}, S\,{\sc vi}, C\,{\sc iii}, N\,{\sc iii}, O\,{\sc vi}, S\,{\sc iv}, 
P\,{\sc iv} \& P\,{\sc v}. The spectrum is also rich in ISM lines: molecular hydrogen, 
oxygen and some metal lines such as C\,{\sc ii}, Ar\,{\sc i}, Fe\,{\sc ii} and many 
P\,{\sc ii} lines. The broad absorption lines can be seen better when 
considering the difference between exposures as a function of the orbital phase 
(due to the inclination of the system). Since only the broad absorption 
lines vary with the phase, a difference graph clearly shows the broad lines. 
In the difference graph the absorption lines appear as broad emission and 
are blue shifted by about 1.5~\AA . The {\it FUSE} spectrum of RW Sex combined
with existing HST and IUE spectra was analyzed in \citet{lin10}.   

\paragraph{SS Aurigae} 
The {\it FUSE} spectrum of the DN SS Aur was taken in quiescence and it is modeled 
with a WD spectrum assuming E(B-V)=0.08. We identify some broad emission
features from C\,{\sc iii} (977) and the oxygen doublet, and identify (with matching
the synthetic spectrum) absorption lines from C\,{\sc ii} (1010, usually seen in lower
temperature WDs),Si\,{\sc iii} (1113, and $\sim$1145), and Si\,{\sc iv} (1123 \& 1128). 
The C\,{\sc iii} (1175) line is not detected at all and the continuum there appear to
be fairly flat. In spite of the low S/N we do identify many ISM molecular hydrogen 
and some metal absorption lines. 
The {\it FUSE} spectrum of this system was processed and modeled in 
\citet{god08b}. 

\paragraph{SS Cygni} 
The spectrum consists of eight individual {\it FUSE} exposures (orbits) combined together. 
The system was in a relatively low state at the time of the observations. 
The {\it FUSE} spectrum is characterized by a relatively low flux with 
broad emission 
lines (Hydrogen Lyman series, C\,{\sc iii}, N\,{\sc iii} and O\,{\sc vi} doublet). 
The spectrum has a ``forest'' of ISM lines: molecular hydrogen (marked in the top of each 
panel in the figure), atomic hydrogen (marked below each panel), 
oxygen (O\,{\sc i}) and few metal lines (marked in the lower part of the panels). 
The spectrum has been dereddened assuming E(B-V)=0.04. 
The main spectral features that belong to the source are the braod emission lines
from C\,{\sc iii} (977 and 1175) and the oxygen doublet. The absorption lines that
are usually seen in WD photosphere (e.g. from S and Si) are not detected at all and the
continuum appear to be rather featureless. 
A composite WD+Disk model has been fit to this spectrum. 
The disk contributes only 12\% of the flux while the WD contributes 
the remaining 88\% of the flux. The increase in flux in the shorter 
wavelengths of {\it FUSE} could be due to broad and strong emission lines of N\,{\sc iv}, 
S\,{\sc vi}, C\,{\sc iii} and the higher order of the Hydrogen Lyman series 
merging together to form a continuum, but it could also be due to the tail of a hotter
source. 
The distance obtained from this fit is 173~pc, in excellent agreement with its
trigonometric parallax distance. 
The complete analysis is given in \citet{sio10}.  

\paragraph{TX Columbae} 
The {\it FUSE} spectrum of the intermediate polar TX Col is rather poor, 
in many respects, resembling the spectrum of MU Cam, both in its 
quality and in its features. The ISM lines have been marked without labeling. 
There seems to be some 
N\,{\sc iv}, 
N\,{\sc iii}, He\,{\sc ii}  and 
O\,{\sc vi} emission, as well as possibly some 
S\,{\sc iv} (1073~\AA ) emission, but little 
C\,{\sc iii}. 
Additional lines have been marked but are not detected with confidence.

\paragraph{UU Aquilae}  
The {\it FUSE} spectrum of the U Gem-type 
dwarf nova UU Aquilae was taken during quiescence. 
The  spectrum is very noisy at the shorter wavelengths. 
The sharp emission lines are geo- and helio-coronal in origin. 
We identify ISM hydrogen molecular absorption features polluting 
the spectrum over almost the entire {\it FUSE} range. 
We note the absence of the C\,{\sc iii} 1175 absorption line. 
The spectrum was modeled using a single WD component. 
Details on the processing and spectral modeling of the {\it FUSE}
spectrum of UU Aql are given in \citet{sio07}.  
In a first model we assumed a 1 solar mass WD ($log(g)= 8.6$) 
with solar composition except for C=0.001 solar. We obtain 
a temperature of 29,000~K and a distance of 380~pc. 
When we assume a canonical mass of 0.6 solar mass, we obtain a 
temperature of 26,000~K and a distance of 450~pc 
(this model is very similar to the 1 solar mass model). 
The apparent discrepancy between the model and the observed 
spectrum shortward of 1040~\AA\ is due mainly to ISM Hydrogen molecular 
absorption. The Si\,{\sc iii} and S\,{\sc iv} lines between 1100~\AA\ and 1150~\AA\ 
are relatively well modeled.  

\paragraph{V3885 Sgr} 
The {\it FUSE} spectrum of the nova-like V3885 Sgr is 
shown in Figure 6 with a WD+disk model fit  
including the contribution from a hot
boundary layer.  The $0.7M_{\odot}$ WD 
model has a temperature of 60,000K, a projected rotational velocity of
200~km~$s^{-1}$, and solar abundances.
The disk model has $\dot{M}=3 \times 10^{-9} M_{\odot}$/yr with
an inclination of 60$^{\circ}$ and includes two inner (boundary layer) rings
with T=175,000K. The WD (dotted line) contributes only 22\% of
the flux and the disk+BL (dashed line) contributes 78\%. 
The sharp absorption lines are from the ISM. All the broad absorption
lines originate from a corona above the disk and/or boundary layer
and are blue shifted by $\sim 1 - 1.5$~\AA\ .  The lines are all marked
at their rest wavelengths to emphasize the shift.  All the absorption
lines are colored in blue and only the continuum (in red) is modeled.

\paragraph{V405 Aurigae}  
The {\it FUSE} spectrum has broad emission lines from highly ionized species 
C\,{\sc iii}, O\,{\sc vi}, S\,{\sc iv}, 
N\,{\sc iii} together with very sharp emission lines possibly mostly from airglow 
contamination (Hydrogen Lyman series, O\,{\sc i} series, 
S\,{\sc vi}). There are many ISM molecular hydrogen abosprtion lines. 
Ar\,{\sc i}, N\,{\sc i}, and Si\,{\sc ii} lines  are also from the ISM. 
The relatively low S/N of the continuum prevents us from identifying more 
features.   

\paragraph{V471 Tauri} 
V471 Tau was observed under the {\it FUSE} program M112 (Periodic Channel 
Coalignment) together with other bright UV sources to check the alignment 
of the {\it FUSE} channels. During the observation the {\it FUSE} telescope was 
performing X and Y scans in the field of view of the target and therefore 
the FUV flux of V471 Tau was collected only a fraction of the time. 
The present {\it FUSE} spectrum was obtained by integrating only during 
the time V471 Tau was on target with a constant FUV flux. 
In the first and second exposures we selected 250s each of good exposure
time, 
in the 3rd - 300s, in the 4th - 800s, and in the 5th and last 
exposure 500s, totaling 2145s of good exposure time. 
The {\it FUSE} spectrum of V471 Tau (see Figure 1) is mainly that of a WD 
atmosphere with Hydrogen, rotating slowly. There is little 
absorption except for the orders of the Lyman series. 
The Si and C lines were fit using very low abundances 
(see Table 6) and with a rotational velocity of 250~km~$s^{-1}$, 
which is about four times larger than expected. 
The broadening of the lines could possibly be due to an unresolved Zeeman splitting 
as suggested by \citet{sio12} who carried out an analysis of the HST/STIS
spectra of V471 Tau.  

\paragraph{V794 Aquilae}  
The {\it FUSE} spectrum of the NL V794 Aql in a high state is shown with a 
disk spectrum. The {\it FUSE} spectrum has been dereddened assuming E(B-V)=0.20 
estimated from the 2200~\AA\ feature in the IUE spectra. Details of the
processing of the {\it FUSE} spectrum and its spectral modeling are given
in \citet{god07}.  The spectrum exhibits many sharp emission lines possibly
all from terrestrial/solar contamination. The spectrum is basically sliced by
ISM molecular hydrogen absorption lines and also includes some ISM metal 
lines such as Ar\,{\sc i}, Si\,{\sc ii}, Fe\,{\sc ii}, P\,{\sc ii}, N\,{\sc i} and S\,{\sc i}. 
There is no obvious broad emission lines and only a comparison with the disk model
shows reveals that there could be some extra flux in the region of the oxygen
doublet and C\,{\sc iii}. 

\paragraph{VW Hydri} 
The {\it FUSE} spectrum of the DN VW Hyi is shown in quiescence with a 
single WD model. We ignore a possible second component 
(see the analysis of \citet{lon07,lon09}). 
The effect of the the Hydrogen quasimolecular opacity is included in the modeling
and they way it affects the spectrum is indicated with the arrows (the dotted/dashed
line shows a model without its inclusion). 
The {\it FUSE} spectrum has been dereddened assuming the published value E(B-V)=0.01. 
There are many more {\it FUSE} exposures available in the MAST archive,
however, the {\it FUSE} spectrum in the catalog was processed and modeled in 
\citet{god08b}. The additional {\it FUSE} spectra of VW Hyi were analyzed in
\citet{lon07,lon09}.  Towards the shorter wavelengths as well as in the Lyman 
$\beta$ region the flux does not go to zero. It could be due to a second emitting
component (emitting a continuum flux). This second component is observed only
during the early stages after outburst. 

\paragraph{WW Ceti} 
The {\it FUSE} spectrum of the DN WW Cet in quiescence is shown 
with a single WD model.            
The effect of the Hydrogen quasimolecular opacity is included 
in the modeling. 
Towards the very short wavelengths one does not expect much flux
due to the hydrogen cut-off. The fact that we do detect flux there is a
sign of a noisy spectrum due to wrong background substraction and/or to
a noisy detector edge. Since the flux in the bottom of the Lyman $\beta$ 
does not go to zero (at least on the left side not affected by oxygen emission), 
a second component could also be present here contributing flux at wavelengths
$< 950$\AA . We ignore here a possible second component 
(the second component is modeled in \citet{god06}).    
We do not identify any feature due to N\,{\sc iv} (923), S\,{\sc vi} (933, 944),
S\,{\sc iv} (1006), C\,{\sc ii} (1010), and Si\,{\sc iii} (1158). 
All the other lines marked on the figure are identified. 

\paragraph{YY Draconis} 
The {\it FUSE} exposures of the IP YY Dra was heavily affected by airglow 
and the spectrum presented here has been processed using the 
``NIGHT'' option in CalFUSE. The resulting spectrum has basically 
no airglow but the S/N is not very high. 
The spectrum is characterized by a rather flat continuum with 
three broad emission features from C\,{\sc iii} (977\AA ), 
O\,{\sc vi} (doublet) and C\,{\sc iii} (1175\AA ).

\clearpage

\clearpage 

\setlength{\hoffset}{-20mm} 
\begin{deluxetable}{llrlccccc}
\tabletypesize{\scriptsize} 
\tablewidth{0pt}
\tablecaption{Adopted System Parameters} 
\tablehead{
System & RA       &  DEC      & Type/ & P     & i     & E(B-V) &  $M_{wd}$      & d    \\  
Name   & hh:mm:ss & deg:mm:ss & Subtype&(hrs) & (deg) &        & ($M_{\odot}$)  & (pc)     
}
\startdata
AE Aqr    &20:40:09.07&-00:52:16.3&Mg   IP     &9.88      &66  & ... &0.63    &102       \\   
AE Ara    &17:41:04.91&-47:03:27.3&Msc  R      &...       &... &... & ...       &   ...       \\   
AG Dra    &16:01:41.01&+66:48:10.1&Msc  R      &...       &...  & ...     &  ...      &  ...        \\   
AM Cas    &02:26:23.45&+71:18:31.4&DN   ZC     &3.96      &18&0.20  &0.55    &350       \\   
AM Her    &18:16:13.33&+49:52:04.2&Mg   P      &3.09425   &50&  ...    &0.76    &79        \\   
AQ Men    &05:07:53.90&-79:51.22.9&DN   UG     &3.40      &...  &...      & ...       &710       \\   
BB Dor    &05:29:28.59&-58:54:46.3&NL   VY     &3.58      &08&0.00  & ...       &   ...       \\   
BV Cen    &13:31:18.60&-54:58:32.0&DN   UG     &14.66830  &53&0.10  &1.20    &     238  \\ 
CH UMa    &10:07:00.77&+67:32:48.7&DN   UG     &8.23642   &21&0.00  &1.35    &     314  \\ 
DQ Her    &18:07:30.12&+45:51:32.7&Mg   IP     &4.65      &90&...      &0.60    &525       \\   
DT Aps    &17:22:47.74&-75:09:56.6&DN   SS     & ...         &...  & ...     &  ...      &710       \\   
EK TrA    &15:14:01.47&-65:05:31.3&DN   SU     &1.5091    &58&0.03  &0.46    & ...        \\   
EM Cyg    &19:38:40.01&+30:30:27.0&DN   ZC     &6.98182   &67&0.05  &1.00    &400       \\   
ES Dra    &15:25:31.68&+62:01:00.2&DN   UG     &4.29      &...  &0.00  & ...       & ...        \\   
EX Hya    &12:52:24.47&-29:14:57.5&Mg   IP     &1.63761   &78& ...     &0.79    &65        \\   
EY Cyg    &19:54:36.77&+32:21:54.7&DN   UG     &11.02377  &16&0.00  &1.26    &  ...        \\   
FO Per    &04:08:34.99&+51:14:48.3&DN   UG     &4.        &...  &  ...    &...        &270       \\   
HP Nor    &16:20:49.58&-54:53:23.0&DN   ZC     &  ...        &...  &...      &...        &  ...        \\   
IX Vel    &08:15:18.97&-49:13:20.7&NL   UX     &4.65425   &57&...      &0.82    &96       \\   
MU Cam    &06:25:16.23&+73:34:38.9&Mg   IP     &4.71864   &...  & ...     & ...       & ...         \\   
MV Lyr    &19:07:16.30&+44:01:08.4&NL   VY     &3.19      &12&0.00  &0.73    &550  \\ 
NSV 10934 &18:40:52.45&-83:43:09.7&DN   SU     &1.74      &...  & ...     &...        &150       \\   
P831 57   &03:34:34.20&-64:00:56.3&MSc         & ...         &...  &...      &0.55    &     115  \\ 
RU Peg    &22:14:02.58&+12:42:11.4&DN   UG     &8.9904    &33&0.00  &1.29    &282      \\   
RW Sex    &10:19:56.62&-08:41:56.1&NL   UX     &5.88168   &34&0.01  &0.8     &150       \\   
SS Aur    &06:13:22.40&+47:44:26.1&DN   UG     &4.3872    &38&0.08  &1.08    &201       \\   
SS Cyg    &21:42:42.70&+43:35:09.5&DN   UG     &6.60312   &51&0.04  &0.81    &166       \\   
TX Col    &05:43:20.27&-41:01:56.1&Mg   IP     &5.718     &...  & ...     &0.54    & ...         \\   
UU Aql    &19:57:18.76&-09:19:20.9&DN   UG     &3.92477   &...  &  ...    & ...       &     380  \\ 
V3885 Sgr &19:47:40.53&-42:00:26.4&NL   UX     &4.97186   &65&0.02  &0.70    & 110      \\ 
V405 Aur  &05:57:59.27&+53:53:45.1&Mg   IP     &4.16      &04& ...     &  ...      & ...         \\   
V471 Tau  &03:50:24.97&+17:14:47.4&Msc  PCV    &12.50839  &79&0.00  &0.83    &47        \\   
V794 Aql  &20:17:34.03&-03:39:50.2&NL   VY     &3.68      &60&0.20  &0.88    & ...         \\   
VW Hyi    &04:09:11.34&-71:17:41.1&DN   SU     &1.783     &60&0.01  &0.86    &65       \\   
WW Cet    &00:11:24.80&-11:28:44.1&DN   UG     &4.22      &54&0.00  &0.85    &200       \\   
YY Dra    &11:43:38.51&+71:41:19.2&Mg   IP     &3.96      &42&      &0.83    &155        \\    
\enddata  
\\ 
\small{
Note: the systems are listed in alphabetical order of the first letters, rather
than of the constellation, since this is the way the systems are returned by the
catalog search page.
} 
\end{deluxetable}

\clearpage

\begin{deluxetable}{llcccc}
\tabletypesize{\scriptsize} 
\tablewidth{0pt}
\tablecaption{{\it FUSE} Observation Log}
\tablehead{
System & DATAID  & Obs. date & Obs. time (UT) &   Exp.time  &  state   \\  
Name   &         & yyyy-mm-dd & hh:mm:ss & sec   &        
}
\startdata
AE Aqr  &B0340101  &2001-06-17&12:26:34&28821  &                           \\
AE Ara  &D1460201  &2004-05-18&17:41:56&9831   &                           \\
AG Dra  &S3120102  &2000-03-16&15:57:00&2387   &                           \\
AM Cas  &G9251402  &2006-10-19&03:17:49&12909  &intermediate                        \\
AM Her  &Z0060101  &2002-05-11&23:39:59&46901  &low                        \\
AQ Men  &G9250201  &2006-11-22&16:16:28&22661  &intermediate                        \\
BB Dor  &H9030301  &2007-07-09&02:51:09&3500   &high                        \\
BV Cen  &D1450301  &2003-04-13&20:26:00&26545  &low     \\  
CH UMa  &D1450201  &2003-04-02&22:00:00&17311  &low    \\  
DQ Her  &D9130501  &2003-06-30&13:20:51&7355   &                           \\
DR Aps  &G9251101  &2006-04-28&10:47:43&68216  &intermediate                        \\
EK TrA  &Z9104301  &2002-06-24&07:22:05&33525  &low                        \\
EK TrA  &Z9104302  &2002-06-24&20:39:39&33842  &low                        \\
EK TrA  &combined  &2002-06-24&07:22:05&67367  &low                        \\
EM Cyg  &C0100101  &2002-09-05&11:34:11&1479   &low                        \\
ES Dra  &G9251601  &2006-11-19&23:08:35&24554  &intermediate                        \\
EX Hya  &A0840102  &2000-05-19&01:19:36&9913   &                           \\
EY Cyg  &D1450101  &2003-07-16&17:35:26&19295  &low                        \\
FO Per  &G9251501  &2007-02-11&03:02:06&2716   &high                        \\
HP Nor  &G9250601  &2007-04-13&17:13:23&3952   &intermediate                        \\
IX Vel  &Q1120101  &2000-04-15&15:34:47&6086   &high                          \\
MU Cam  &E9890801  &2004-11-06&06:47:54&13909  &                           \\
MV Lyr  &C0410301  &2002-07-07&11:56:39&11209  &low    \\  
NSV10934&G9251201  &2006-06-28&14:08:26&10844  &low                        \\
NSV10934&G9251202  &2006-06-30&17:46:04&14177  &low                        \\
NSV10934&combined  &2006-06-28&14:08:26&25021  &low                        \\
P83157  &D9131401  &2003-11-03&02:26:19&11865  &       \\  
RU Peg  &C1100101  &2002-07-04&17:10:03&3026   &low                        \\
RW Sex  &B1040101  &2001-05-13&14:48:05&25614  &high                        \\
SS Aur  &C1100201  &2002-02-13&06:59:53&12733  &low                        \\
SS Cyg  &P2420101  &2001-09-04&08:22:31&18583  &low                        \\
RX Col  &D9050201  &2003-12-31&07:00:22&3330   &                           \\
UU Aql  &C1100301  &2004-05-16&13:48:00&16121  &low    \\  
V3885 Sgr&P1870101 &2000-05-24&01:51:55&12392  &high    \\ 
V405 Aur&D0800101  &2003-10-01&02:06:07&21581  &                           \\
V471 Tau&M1124002  &2001-01-12&05:27:23&2145   &                           \\
V794 Aql&D1440101  &2004-05-13&19:50:22&13438  &high                        \\
VW Hyi  &B0700201  &2001-08-18&10:29:41&17435  &low                        \\
WW Cet  &D1450401  &2003-07-26&05:30:03&16291  &low                        \\
YY Dra  &C0410201  &2002-01-30&21:29:16&15139  &                           \\
\enddata
\\ 
\small{
Note: the systems are listed in alphabetical order of the first letters, rather
than of the constellation, since this is the way the systems are returned by the
catalog search page.
} 
\end{deluxetable}

\clearpage

\setlength{\voffset}{-25mm} 
\setlength{\hoffset}{-10mm} 
\begin{deluxetable}{ll}
\tabletypesize{\scriptsize} 
\tablewidth{0pt}
\tablecaption{Lines in FUSE Spectra of Cataclysmic Variables and Related Objects} 
\tablehead{
Ion                     & $\lambda$(\AA)    
}
\startdata
He\,{\sc ii} (Balmer n=20) & 920.56   \\  
N\,{\sc iv}                & 921.46, 921.99, 922.52, 923.06, 923.22, 923.68, 924.28, 924.91   \\  
He\,{\sc ii} (Balmer n=18) & 922.75   \\  
He\,{\sc ii} (Balmer n=17) & 924.15   \\  
He\,{\sc ii} (Balmer n=15) & 927.85   \\  
He\,{\sc ii} (Balmer n=14) & 930.34   \\  
He\,{\sc ii} (Balmer n=13) & 933.45   \\  
S\,{\sc vi}                & 933.38   \\ 
He\,{\sc ii} (Balmer n=12) & 937.39   \\  
He\,{\sc ii} (Balmer n=11) & 942.51   \\  
S\,{\sc vi}                & 944.52   \\ 
He\,{\sc ii} (Balmer n=10) & 949.33   \\  
P\,{\sc iv}                & 950.66     \\ 
He\,{\sc ii} (Balmer n=9)  & 958.70   \\  
P\,{\sc ii}                & 961.04, 962.12, 962.57, 963.62, 963.80, 964.95, 965.40   \\  
Ne\,{\sc vii}              & 973.35   \\  
C\,{\sc iii}               & 977.02   \\  
N\,{\sc iii}               & 989.79 \& 991.56   \\  
He\,{\sc ii} (Balmer n=7)  & 992.36   \\  
Ne\,{\sc vi}               & 992.73   \\  
Ne\,{\sc vi}               & 997.17   \\  
Ne\,{\sc vi}               & 999.29   \\  
Ne\,{\sc vi}               & 1005.79   \\  
S\,{\sc iv}                & 1006.39    \\ 
C\,{\sc ii}                & 1009.86, 1010.08,1010.37    \\ 
Ne\,{\sc vi}               & 1010.32   \\  
S\,{\sc iii}               & 1015.50, 1015.57, 1015.78   \\  
S\,{\sc iii}               & 1021.11, 1021.32   \\  
O\,{\sc vi}                & 1031.93   \\  
O\,{\sc vi}                & 1037.62   \\  
S\,{\sc iv}                & 1062.66    \\ 
Si\,{\sc iv}               & 1066.61, 1066.64, 1066.65    \\ 
Fe\,{\sc iii}              & 1066.20    \\ 
S\,{\sc iv}                & 1072.98 \& 1073.52   \\ 
He\,{\sc ii} (Balmer n=5)  & 1084.94   \\  
S\,{\sc iv}                & 1098.36, 1098.93, 1099.48, 1100.05    \\ 
Si\,{\sc iii}              & 1108.36, 1109.94, 1113.2     \\ 
P\,{\sc v}                 & 1117.98    \\ 
P\,{\sc iv}                & 1118.55    \\ 
S\,{\sc ii}                & 1124.40, 1124.99    \\ 
P\,{\sc v}                 & 1128.01    \\ 
Si\,{\sc iv}               & 1122.49    \\ 
Si\,{\sc iv}               & 1128.33    \\ 
Ne\,{\sc v}                & 1136.53   \\  
Fe\,{\sc iii}              & 1142.96    \\ 
Si\,{\sc iii}              & 1140.55, 1141.58,1142.29, 1144.31, 1144.96, 1145.67    \\ 
Ne\,{\sc v}                & 1145.62   \\  
Si\,{\sc iii}              & 1155.00, 1155.96, 1156.78, 1158.10, 1160.25, 1161.58    \\ 
C\,{\sc iii}               & 1174.61, 1174.93, 1175.26, 1175.59, 1175.71, 1175.99, 1176.10, 1176.37, 1176.77  
\enddata
\end{deluxetable}

\clearpage

\setlength{\hoffset}{-20mm} 
\begin{deluxetable}{lccccccccccccccccc}
\tabletypesize{\scriptsize} 
\tablewidth{0pt}
\tablecaption{Main Line Identifications} 
\tablehead{
System&N\,{\sc iv}&S\,{\sc vi}&C\,{\sc iii}&N\,{\sc iii}+He\,{\sc ii}&S\,{\sc iv}&C\,{\sc ii}&O\,{\sc vi}&S\,{\sc iv}&Si\,{\sc iv}&He\,{\sc ii}&S\,{\sc iv}&Si\,{\sc iii}&P\,{\sc v}&Si\,{\sc iv}&P\,{\sc v}+Si\,{\sc iv}&Si\,{\sc iii}&C\,{\sc iii}\\
Name  &      923  &  933      &  977        &  990                   &  1006    &  1010     &   1032    &   1063    & 1066       &  1085      &  1100     &      1110   & 1118     & 1123       & 1128                  &  1143       &  1175       \\ 
           &              &  944      &                &                           &              &              &  1038      &   1073    &                &               &              &               &            &               &                          &  1158       &       
}
\startdata
AE Aqr & A        & D,D       & C                &  B                     &   D      &  D         & A,A         & C,B       &  D         &   A        &   D       &   B         & D        & D          &  D                    &  D,D        &     C   \\  
AE Ara & C        & C,C       & C               &  D                     &   C      &  D         & B,B         & D,B       &  D         &   C        &   D       &   D         & D        & D          &  D                    &  D,D        &     B   \\    
AG Dra & D        & B,A       & A               &  C,A                   &   D      &  D         & A,A       & C,A       &  D         &   A        &   D       &   D         & D        & D          &  D                    &  D,D        &     D    \\ 
AM Cas & C        & C,C       & C              &  C,C                   &   D      &  D        & C,D       & D,D       &  D         &   C        &   C       &   C         & D        & C          &  C                    &  D,D        &     D    \\   
AM Her & D        & D,D       & B              &  D                     &   D        &  D         & C,C       &   D,D     & D          &   C        &  D        &     D       & D        & D          & D                     &  D,D        &     D     \\ 
AQ Men & D        & D,D       & D             & C                      &   D        &   D        & D,D       & C,C       & D          &   D        &  C        &     C       & C        & C          & C                     &  C,C        &     C      \\ 
BB Dor & D        & C,C       & C              & D                      &   D        &  C         & A,A        & B,B       & C          &   D        &  D        &   B         & D        & A          & A                     &  C,D        &     B     \\  
BV Cen & D        & C,D       &  B             &  D                     &   D      &   C         & C,D        & D,C       & D          &  D         &  D        &   C         & D        & A          & A                     &  D,D        &     C    \\ 
CH UMa &     C    & D,C       &  C          &  D                     &  D       &   D          &  A,B       & D,C       & D          &  C         &  D        &     D       & C        &  D         &  D                    &   D         &   A      \\ 
DQ Her &    B     &   B,B       &   A          &  B,B                   &   D      &  D        &  A,A      &   A,A     &   D        &   A        &  C        &       C     &  B       & D          &  B                    &  D,D        &  A      \\  
DT Aps &      D   &   C       &   C              &  C                     &   D      &   D       &  D,D      &   C,D     &  D         &   D        &    D      &       C     &  D       &  C         &  C                    &  D,D        &   C     \\  
EK TrA &      D   &   D       &   C               &   C                    &   D      &   D       &  C,C      &   C,D     & C          &   D        &   D       &   B         &  D       &  B         & D,B                   &  D,D        &   C      \\  
EM Cyg &      D   &  D,C      &  C          &  D                     &   D      &   D       &  B,B      &   C,A     & D          &  D         &  C        &      B      & D        &  B         &  B                    &  D,D        &  C       \\ 
ES Dra &      D   &  C,C      &  C          &  C                     &  B       &  B        &  C,C      &   D,D     & C          &    D       &   B       &       B     &  D       &  B         & B                     &  C,D        &   C     \\  
EX Hya &      D   &  A,A      &  A          &  A,A                   &   D      &  A        &  A,A      &   A,A     & A          &   C        &   D       &       A     &  D       & A          &  B                    &  D,D        &   A      \\  
EY Cyg &     D    &  D,D      &   D         &   C                    &   D      &   D       &  C,C      &   D,C     &  D         &   D        &  D        &       C     &  D       & D          & C                     &  C,D        &   D      \\ 
FO Per &      D   &  C,C      &  B          &   D                    &   D      &   D       &  C,C      &   D,C     & C          &   D        &   D       &       B     &  C       & C          & C                     &  C,D        &  B       \\ 
HP Nor &    D     &  C,C      &   C         &  C                     &    D     &    D      &  D,D      &   D,C     &  D         &   D        &   D       &       C     &  D       &  C         &  D                    &  D,D        &   C      \\ 
IX Vel &    A     &  A,A      &    A        &  B                     &    D     &    D      &  B,B      &   A,A     &  D         &   D        &   D       &      D      &  B       &  D         &   B                   &  D,D        &     A    \\ 
MU Cam &      C   &   C,C     &   C         &   C                    &    D     &   C       &    B,C    &   C,D     &   D        &    D       &   D       &       D     & D        &  D         &  D                    &  D,D        &   D      \\ 
MV Lyr &      D   & D,D       &    C        &   C                    &   D      &    C      &  D,D      &   B,B     &   A        &   B        &  D        &       B     &   C      &  A         &  A                    &  D,D        &     A    \\ 
NSV 10934 &   D   &  C,C      &     C       &  D                     &   D      &    D      &  B,B      &    D,D    &  D         &    D       &   D       &       D     &   D      &  D         &   D                   &  D,D        &    B     \\ 
P831 57 &    D    &  D,D      &    B        &  B                     &   C      &   B       &    D,D    &   A,A     &  A         &   B        &    D      &     A       &     D    &    A       &  A                    &   D,D       &   A      \\ 
RU Peg &  C       & C,C       &   C         &  B,B                   &   C      &  C        &  B,B      & C,A       & A          &  C         &  C        &     C       & D        & D          &  C                    &  D,D        &  B       \\ 
RW Sex &     B    & B,B       &  B          &  B                     &  C       &  C        &   D,D     &   D,B     & D          &  C         &  D        &   D         &  B       &   B        &  B                    &  C,D        &  A       \\ 
SS Aur &     D    & D,D       &  B          &  C                     &  D       &   B       &  C,C      &   D,C     &   B        &  D         &  D        &      C      &  D       &    C       &   D,C                 &  C,C        &   D      \\ 
SS Cyg &    D     &  D,D      &  A          &  C                     &  D       &   D       &  B,B      &   D,D     &  D         &   D        &   D       &   D         &   D      &  D         &  D                    &  D,D        &  A       \\ 
TX Col &     C    &  D,D      &  D          &   C                    &  D       &  D        &  B,B      &   C,C     &  D         &   C        &  D        &      D      &    D     &    D       &   D                   &  D,D        &  C       \\ 
UU Aql &   D      & C,C       &  C          & C,C                    &   D      &  D        &  C,C      &   D,D     & D          &   D        &    D      &        C    &   D      &  C         &  C                    &  D,C        &  D       \\ 
V3885 Sgr &  B    & B,B       &   B         &  C                     &  D       &  D        &  B,B      &   A,A     &  D         &  D         &  D        &     B       &   A      &  C         & B                     &  D,D        &    A     \\ 
V405 Aur &    D   & B,B       & A           &  C                     &  D       &   D       &  A,C      &   D,D     &  D         &   B        &  D        &      D      &  D       &  D         &   D                   &  D,D        &   B      \\ 
V471 Tau &   D    & D,D       &  D          &  D                     &  D       &   D       &  D,D      &   D,D     &  D         &  C         &   D       &    C        &   D      & D          &  C                    &  D,D        &    C     \\ 
V794 Aql &   C    & C,C       &  C          &   C                    &  D       &   D       & C,C       &  D,D      &  D         &   C        &  D        &    D        &    D     &   D        &   D                   &  D,D        &     D    \\ 
VW Hyi &     D    &  D,D      &  C          &  C                     &  D       &  D        &  C,C      &   D,D     &  C         &   B        &   C       &   A         &  D       &   C        &  A                    &  B,D        &  A       \\ 
WW Cet &     D    & D,D       & C           &   D                    &  D       &   D       &  C,C      &  D,D      & D          &  C         &   C       &   B         &   D      &    D       &   C                   &  C,D        &   C      \\ 
YY Dra &     D    &   D,D     &  C          &  D                     &  D       &   D       &  C,C      &   D,D     &  D         &   D        &   D       &      D     &   D       &  D         &  D                    &  D,D        &   C     \\ 
\enddata
\small{ 
Note: 
{\bf A} denotes a line that has been correctly identified with no contamination 
and a good S/N. 
{\bf B} denotes a line that has been correctly identified but contaminated,
blended, and/or with a lower S/N. 
{\bf C} denotes a possible line identification but uncertain due
to low S/N. 
{\bf D} denotes a line that has not been identified and the marking on 
the graph is for information only, or for comparison with the theoretical
spectrum and cannot be relied on.  
}     
\end{deluxetable}

\clearpage

\setlength{\voffset}{-00mm} 
\setlength{\hoffset}{-10mm} 
\begin{deluxetable}{lrllrllrllrllrllr}
\tabletypesize{\scriptsize} 
\tablewidth{0pt}
\tablecaption{ISM lines in FUSE Spectra} 
\tablehead{
H\,{\sc i}  & $\lambda$(\AA) & $|$ & 
O\,{\sc i}  & $\lambda$(\AA) & $|$ & 
H$_2$       & $\lambda$(\AA) & $|$ &  
H$_2$       & $\lambda$(\AA) & $|$ &  
H$_2$       & $\lambda$(\AA) & $|$ & 
Ion         & $\lambda$(\AA)  
}
\startdata
            30u & 912.70  & $|$ &   
            20u & 921.86  & $|$ &   
L16R2           & 933.24  & $|$ & 
L10P3           & 987.77  & $|$ & 
L5P1            & 1038.16 & $|$ & 
N\,{\sc i}      & 953.42      \\  

            29u & 912.77  & $|$ &   
            19u & 922.20  & $|$ &   
W4Q3            & 933.58  & $|$ & 
W1R2            & 987.97  & $|$ & 
L5R2            & 1038.69 & $|$ & 
N\,{\sc i}      & 953.66      \\   

            28u & 912.84  & $|$ &   
            18u & 924.95  & $|$ & 
L15R0           & 938.47  & $|$ & 
L9R0            & 991.38  & $|$ & 
L5P2            & 1040.37 & $|$ & 
N\,{\sc i}      & 953.97      \\  

            27u & 912.92  & $|$ &   
            17u & 925.45  & $|$ & 
L15R1           & 939.12  & $|$ & 
L9R1            & 992.01  & $|$ & 
L5R3            & 1041.16 & $|$ & 
N\,{\sc i}      & 954.10      \\  

            26u & 913.01  & $|$ &   
            16u & 929.52  & $|$ & 
L15P1           & 939.71  & $|$ & 
L9P1            & 992.81  & $|$ & 
L5P3            & 1043.50 & $|$ & 
P\,{\sc ii}     & 961.04      \\  

            25u & 913.10  & $|$ &   
            15u & 930.26  & $|$ & 
W3Q3            & 950.40  & $|$ & 
L9R2            & 993.55  & $|$ & 
L4R0            & 1049.37 & $|$ & 
P\,{\sc ii}     & 963.80      \\  

            24u & 913.22  & $|$ &   
            14u & 936.63  & $|$ & 
L13R1           & 955.06  & $|$ & 
L9P2            & 994.87  & $|$ & 
L4R1            & 1049.96 & $|$ & 
N\,{\sc i}      & 964.63    \\  

            23u & 913.34  & $|$ &   
            13u & 937.84  & $|$ & 
L13P1           & 955.71  & $|$ & 
L9R3            & 995.97  & $|$ & 
L4P1            & 1051.03 & $|$ & 
C\,{\sc iii}    & 977.02      \\  

            22u & 913.48  & $|$ &   
            12u & 948.69  & $|$ &   
L13R2           & 956.58  & $|$ & 
L9P3            & 997.82  & $|$ & 
L4R2            & 1051.50 & $|$ & 
Si\,{\sc ii}/N\,{\sc iii} & 989.80 \\ 

            21u & 913.64  & $|$ &   
            11u & 950.88  & $|$ &   
L13P2           & 957.65  & $|$ & 
L8R0            & 1001.82 & $|$ & 
L4P2            & 1053.28 & $|$ & 
S\,{\sc iii}    & 1012.50     \\  

            20u & 913.83  & $|$ &  
            10u & 971.74  & $|$ &   
L13R3           & 958.95  & $|$ & 
L8R1            & 1002.45 & $|$ & 
L4R3            & 1053.98 & $|$ & 
Si\,{\sc ii}    & 1020.70     \\  

            19u & 914.04  & $|$ & 
                & 973.23  & $|$ & 
L12R0           & 962.98  & $|$ & 
L8P1            & 1003.29 & $|$ & 
L4P3            & 1056.47 & $|$ & 
C\,{\sc ii}     & 1036.34     \\  

            18u & 914.29  & $|$ &  
                & 973.89  & $|$ &   
L12R1           & 963.61  & $|$ & 
L8R2            & 1003.98 & $|$ & 
L3R0            & 1062.88 & $|$ & 
C\,{\sc ii}     & 1037.02     \\  

            17u & 914.58  & $|$ &  
            9u  & 972.14  & $|$ &   
W2R2            & 965.79  & $|$ & 
L8P2            & 1005.39 & $|$ & 
L3R1            & 1063.46 & $|$ & 
Ar\,{\sc i}     & 1048.20     \\  

            16u & 914.92  & $|$ &  
                & 973.64  & $|$ &   
W2Q1            & 966.09  & $|$ & 
L8R3            & 1006.41 & $|$ & 
L3P1            & 1064.61 & $|$ & 
Fe\,{\sc ii}    & 1055.26     \\  

            15u & 915.33  & $|$ & 
                & 974.29  & $|$ &   
W2R3            & 966.78  & $|$ & 
W0R0+1          & 1008.50 & $|$ & 
L3R2            & 1065.00 & $|$ & 
Fe\,{\sc ii}    & 1063.18    \\ 

            14u & 915.82  & $|$ &  
            7u  & 976.45  & $|$ &   
W2Q2            & 967.28  & $|$ & 
W0R2            & 1009.02  & $|$ & 
L3P2            & 1066.90 & $|$ & 
Ar\,{\sc i}     & 1066.66     \\  

            13u & 916.43  & $|$ &  
                & 977.96  & $|$ & 
L12R3           & 967.67  & $|$ & 
W0R1            & 1009.77 & $|$ & 
L3R3            & 1067.48 & $|$ & 
Fe\,{\sc ii}    & 1081.88    \\ 

            12u & 917.18  & $|$ & 
                & 978.62  & $|$ &
W2P2            & 968.29  & $|$ & 
W0R3            & 1010.13 & $|$ & 
L3P3            & 1070.14 & $|$ & 
Fe\,{\sc ii}    & 1083.42    \\ 

            11u & 918.13  & $|$ & 
            5u  & 988.58  & $|$ &   
W2Q3            & 969.05  & $|$ & 
W0R2            & 1010.94 & $|$ & 
L2R0            & 1077.14 & $|$ & 
N\,{\sc ii}     & 1083.99     \\  

            10u & 919.35  & $|$ &  
                & 988.65  & $|$ &   
W2P3            & 970.56  & $|$ & 
W0P2            & 1012.17 & $|$ & 
L2R1            & 1077.70 & $|$ & 
Fe\,{\sc ii}    & 1096.88     \\  

           Ly$\iota$  & 920.96  & $|$ &  
                & 988.77  & $|$ & 
L11P1           & 973.34 & $|$ & 
L7R1            & 1013.44 & $|$ & 
L2P1            & 1078.93 & $|$ & 
Fe\,{\sc ii}    & 1112.05     \\  

           Ly$\theta$ & 923.15  & $|$ &  
                & 990.13  & $|$ & 
L11R2           & 974.16 & $|$ & 
L7P1            & 1014.35 & $|$ & 
L2R2            & 1079.23 & $|$ & 
Fe\,{\sc ii}    & 1121.98     \\  

           Ly$\eta$   & 926.23  & $|$ &   
                & 990.80  & $|$ & 
L11P2           & 975.34 & $|$ & 
L7R2            & 1014.97 & $|$ & 
L1R0            & 1092.20 & $|$ & 
Fe\,{\sc ii}    & 1122.53     \\  

           Ly$\zeta$  & 930.75  & $|$ & 
            4u  & 1025.76 & $|$ & 
L10R0           & 981.44 & $|$ & 
L7P2            & 1016.46 & $|$ & 
L1R1            & 1092.73 & $|$ & 
Fe\,{\sc ii}    & 1125.45     \\  

           Ly$\epsilon$ & 937.80  & $|$ &  
                & 1027.43 & $|$ & 
L10R1           & 982.07 & $|$ & 
L7R3            & 1017.42 & $|$ & 
L1P1            & 1094.05 & $|$ & 
Fe\,{\sc ii}    & 1127.10     \\  

           Ly$\delta$ & 949.74  & $|$ &  
                & 1028.16 & $|$ &  
L10P1           & 982.84  & $|$ & 
L7P3            & 1019.50 & $|$ & 
L1P3            & 1099.79 & $|$ & 
Fe\,{\sc ii}    & 1133.67     \\  

           Ly$\gamma$ & 972.54  & $|$ & 
            3u  & 1039.23 & $|$ & 
L10R2           & 983.59 & $|$ & 
L6R0            & 1024.37 & $|$ & 
L0R0            & 1108.13 & $|$ & 
N\,{\sc i}      & 1134.17       \\       

           Ly$\beta$  & 975.00  & $|$ & 
                & 1040.94 & $|$ & 
L10P2           & 984.86 & $|$ & 
L6R1            & 1024.99 & $|$ & 
L0R1            & 1108.63 & $|$ & 
N\,{\sc i}      & 1134.42       \\       

                          &         & $|$ &   
                & 1041.69 & $|$ &   
W1R0+1          & 986.60 & $|$&   
L6R2            & 1026.53 & $|$ & 
L0P1            & 1110.06 & $|$ & 
N\,{\sc i}      & 1134.98       \\       

                &     & $|$ &   
                &     & $|$ &   
W1R2            & 986.24 & $|$ &   
L6P2            & 1028.10 & $|$ & 
L0R2            & 1110.12 & $|$ & 
Fe\,{\sc ii}    & 1142.34      \\        

                &     & $|$ &   
                &     & $|$ &   
W1R1            & 986.80  & $|$ &   
L5R0            & 1036.55 & $|$ & 
                &         & $|$ & 
Fe\,{\sc ii}    & 1143.23      \\        

                &     & $|$ &   
                &     & $|$ &   
W1R3            & 987.45   & $|$ &   
L5R1            & 1037.15 & $|$ & 
                &         & $|$ & 
Fe\,{\sc ii}    & 1144.94      \\          

                &     & $|$ &   
                &     & $|$ &   
                &     & $|$ &   
                &     & $|$ &   
                &     & $|$ &   
P\,{\sc ii}     & 1152.82      \\          
\enddata 
\\ 
The ISM molecular hydrogen lines are identified by their band 
(Werner/W or Lyman/L),, upper vibrational level (1-13), and rotational 
transition (R, P, or Q with lower rotational state J=1-3). 
\end{deluxetable}

\clearpage 

\setlength{\voffset}{-00mm} 
\setlength{\hoffset}{-23mm} 
\begin{deluxetable}{lccccccccccccl}
\tabletypesize{\scriptsize} 
\tablewidth{0pt}
\tablecaption{Accretion Disk \& White Dwarf Synthetic Spectral Model Fits}
\tablehead{
System  &i &$E_{B-V}$&$log(g)$& $T_{wd}$ &$V_{rot} \sin{i}$&C            & Si           & S           & N           & d  &$log(\dot{M})$ &model & Reference   \\
Name    & deg &      &  cgs    & $10^3$K &  km~$s^{-1}$           & C$_{\odot}$ & Si$_{\odot}$ & S$_{\odot}$ & N$_{\odot}$ & pc & $M_{\odot}$/yr &      &           
}
\startdata
AM Cas  &18&0.00  &8.25  &36   &500 &1.00  &1.0  &1.0&1.0 &373 &-9.7     &wd+disk  &  \citet{god09a} \\     
        &18&0.20  &7.75  &35   &400 &1.00  &1.0  &1.0&1.0 &331 &-8.5     &wd+disk  &     "~~~~~"     \\ 
BB Dor  &...  &0.00  &8.30  &37   &400 &1.00  &3.0  &20.&1.0 &211 & ...     &wd       &  \citet{god08a} \\  
        & 8&0.00  &8.30  &...  &... &1.00  &3.0  &20.&1.0 &665 &-9.0     &disk     &  "~~~~~"     \\  
        & 8&0.00  &8.30  &32   &400 &1.00  &3.0  &20.&1.0 &700 &-9.0     &wd+disk  &  "~~~~~"     \\  
        &80&0.00  &8.30  &37   &400 &1.00  &3.0  &20.&1.0 &246 &-9.0     &wd+disk  &  "~~~~~"     \\  
BV Cen  &...  &0.10  &9.00  &40   &400 &1.0   &1.0  &1.0&1.0 &238 &...      &wd       &  \citet{sio07}  \\  
CH UMa  &...  &0.00  &9.50  &40   &200 &$<$1.0&$<$1.0&1.0&1.0&314 &...      &wd       &  \citet{sio07}  \\ 
EK TrA  &...  &0.03  &8.00  &17   &200 &0.10  &0.6  &1.0&1.0 &126 &...      &wd       &  \citet{god08b} \\            
EM Cyg  &60&0.00  &8.60  &50   &100 &1.00  &30.0 &10.&10. &382 &-10.0    &wd+disk  &  \citet{god09a} \\     
ES Dra  &...  &0.00  &7.8   &35   &700 &1.0   &10.0 &50.&1.0 &770 &...      &wd       &  \citet{god09a} \\      
        & 5&0.00  &9.00  &...  &... &1.0   &10.0 &50.&1.0 &1754&-9.5     &disk     &   "~~~~~"     \\  
EY Cyg  &...  &0.00  &9.00  &30   &100 &0.03  &0.6  &6.0&2.0 &530 &...      &wd       &  \citet{god08b}  \\  
FO Per  &75&0.00  &7.50  &21   &200 &1.00  &1.0  &1.0&1.0 &291 &-8.5     &wd+disk  &  \citet{god09a}  \\
        &18&0.30  &7.50  &40   &200 &1.00  &1.0  &1.0&1.0 &254 &-8.5     &wd+disk  &   "~~~~~"     \\  
HP Nor  &18&0.20  &8.25  &...  &... &1.00  &1.0  &1.0&1.0 &1800&-8.0     &disk     &  \citet{god09a}  \\ 
MV Lyr  &12&0.00  &8.01  &45   &200 &1.00  &1.0  &1.0&1.0 &550 &...      &wd       &  \citet{god11}, this work \\
P83157  &...  &0.00  &7.85  &29   &001 &0.01  &0.02 &0.01&0.01&115 & ...    &wd       &  \citet{bar12}   \\     
RU Peg  &...  &0.00  &8.80  &70   &40  &0.20  &0.2  &10.&$>$1.0&282 &...      &wd     &  \citet{god08b}  \\ 
SS Aur  &...  &0.08  &8.93  &34   &400 &1.00  &1.0  &1.0&1.0 &200 & ...     &wd       &  \citet{god08b}  \\
SS Cyg  &50&0.04  &8.30  &46   &200 &1.0   &1.0  &1.0&1.0 &173 &-10.0    &wd+disk  &  \citet{sio10}   \\      
UU Aql  &...  &0.00  &8.60  &29   &200 &0.001 &1.0  &1.0&1.0 &380 & ...     &wd       &  \citet{sio07}   \\     
V3885 Sgr&60&0.02 &8.00  &60   &200 &1.0   &1.0  &1.0&1.0 &110 &-8.5     &wd+disk  &  \citet{lin09}, this work    \\ 
V471 Tau&79&0.00  &8.30  &33.4 &250 &0.0003&0.015&0.0&0.0 &47  & ...     &wd       &  \citet{sio12}, this work \\
V794 Aql&60&0.20  &8.60  &...  &... &1.00  &1.0  &1.0&1.0 &585 &-8.5     &disk     &  \citet{god07}    \\   
VW Hyi  &...  &0.01  &8.00  &22   &400 &0.25  &1.8  &1.0&3.0 &60  & ...     &wd       &  \citet{god08b}   \\   
WW Cet  &...  &0.00  &8.35  &27   &600 &0.10  &0.3  &1.0&2.0 &213 & ...     &wd       &  \citet{god06}    \\    
\enddata
\end{deluxetable}

\clearpage

\setlength{\hoffset}{-00mm} 
\begin{deluxetable}{ll}
\tabletypesize{\scriptsize} 
\tablewidth{0pt}
\tablecaption{References}
\tablehead{
System           & References    \\ 
Name             &       
}
\startdata
AE Aqr           &  see \citet{muk12} and references therein \\ 
AE Ara       &            \citet{can33,all74,all78,all80,sch89}   \\    
             &            \citet{mik89,mik97,mik99,mik03}   \\ 
             &            \citet{bel00,sko07,fek10}                      \\   
AG Dra           &  \citet{fek03,gal99,gon99,gre97}                        \\
                 &  \citet{mei79,mik95,mur91,sko05}                         \\
                 &  \citet{sko09,smi96,tom00,tom02}                         \\
                 &  \citet{sch99,vio83,vio84,you05}                        \\ 
AM Cas           &  \citet{hof28,ric61,not84,ric88}                        \\
                 &  \citet{tay96,god09a}                                   \\ 
AM Her           &  see \citet{gan06} and references therein               \\ 
AQ Men           &  \citet{che01,pat02,god09a}                             \\ 
BB Dor           &  \citet{che01,pat05a,pat05b,god08a}                     \\  
BV Cen           &  \citet{vog80,gil82,men86,wil88}                        \\
                 &  \citet{sio07,wat07a,wat07b}                           \\  
CH UMa           &  \citet{bec82,szk84,tho86,fri90}                        \\  
                 &  \citet{sim00,sio07}                             \\  
DQ Her           &  see \citet{muk12} and references therein \\  
DT Aps           &       \citet{vog82,god09a}   \\  
EK TrA          &      \citet{vog80s,has85,war87,gan97,gan01}       \\
                &      \citet{god08b}       \\   
EM Cyg          &      \citet{bur53,kra64,bey67,mum69}        \\ 
                &      \citet{pri75,rob74,ben75,bra77}         \\ 
                &      \citet{mum79,jam81,mum80,bay81}         \\ 
                &      \citet{sto81,will83,beu84,her87}         \\ 
                &      \citet{hac87,zha89,and95,and96}         \\ 
                &      \citet{nor00a,nor00b,nor02,win03,spo03a,spo03b,spo05} \\ 
                &      \citet{wel05,wel07,urb06,ham07} \\
                &      \citet{csi08,god09a}         \\ 
ES Dra          &      \citet{gre86,and91,and03,rin94}  \\  
                &      \citet{god09a}            \\ 
EX Hya          &   see \citet{muk12} and references therein      \\ 
EY Cyg          &       \citet{boc85,sar95,smi97,cos98}      \\
                &       \citet{win01,gan03,tov02,cos04}        \\        
                &       \citet{sio04b,god08b}                  \\      
FO Per          &      \citet{how76,ges78,bru87}       \\ 
                &      \citet{bru89,she07,god09a}      \\         
HP Nor          &         \citet{vog82,pre06,god09a}         \\  
IX Vel          &             \citet{gar82,gar84,war83,war84,egg84}  \\
                &             \citet{war85,sio85,hau87,hau88,beu90}   \\ 
                &      \citet{pol90,sha91,mau91,lon94}     \\
                &      \citet{pri95,van95,kub99,har02}   \\    
                &      \citet{lin07}   \\    
MU Cam          & see \citet{muk12} and references therein      \\ 
MV Lyr          &    \citet{sch81,rob81,sch82}     \\ 
                &    \citet{szk82,ski95,hoa04}      \\    
                &    \citet{lin05,god11}              \\   
NSV 10934       &    \citet{kat02,kat04,god09a}   \\   
P831-57         &    \citet{rod93,rod94,zac04,bar12}   \\   
RU Peg          &    \citet{sto81b,wad82,sha83,fri90,sio02}    \\ 
                &    \citet{joh03,sio04a,urb06,god08b}     \\   
RW Sex          &    \citet{gre82,pat84,hau85}       \\ 
                &    \citet{pol86,mau88,wad88,sha91}       \\ 
                &    \citet{beu92,per97,las01,pri03}           \\
                &    \citet{lin10,bol87,sto00}         \\   
SS Aur          &    \citet{sha83,sha86,har99,lak01} \\ 
                &    \citet{sio04a,god08b,sio08}     \\          
SS Cyg          &    \citet{sha83,hol88a,hol88b,can92,can93} \\    
                &    \citet{har00,les03,lon05,bit07}  \\
                &    \citet{sio10}  \\     
TX Col          & see \citet{muk12} and references therein      \\ 
UU Aql          &       \citet{oco32,isl76,bat79,dav81}          \\ 
                &       \citet{sha87,bat90,stu01}                \\    
                &       \citet{har05,sio07}        \\            
V3885 Sgr       &       \citet{gui82,hau85b}                  \\  
                &       \citet{met89,woo92,per97}                    \\  
                &       \citet{pan00,har02,hart05,kni06,kni07}       \\ 
                &       \citet{rib07,lin09}                          \\  
V405 Aur        & see \citet{muk12} and references therein      \\ 
V471 Tau        &          \citet{you72,ces76,iba78}     \\ 
                &        \citet{you83,gui84,ski88}    \\
                &        \citet{boi88,cle92,bar92}    \\ 
               &        \citet{ram95,vana95,pro96}    \\ 
               &        \citet{dup97,bar97,wer97}     \\ 
               &        \citet{whe98,sio98,obr01,bon01}          \\       
             &          \citet{wal03,wal04,iba05,jef10}     \\
             &          \citet{sio12}      \\       
V794 Aql         &         \citet{szk81,szk88,war82,hon85}  \\
                 &         \citet{hon94,hon98,god07}    \\    
VW Hyi          &     \citet{scho81,mat84,van87a,van87b}   \\ 
                &     \citet{war87,pol90,bel91,mau91e}             \\      
                &     \citet{sio95a,sio95b,sio96,hua96a,hua96b,lon96}             \\ 
                 &         \citet{mau96,whe96,gan96}            \\        
               &        \citet{sio97,sio01,pan03,lon07}       \\      
               &        \citet{god04,god05,god08b,lon09}    \\  
WW Cet       &            \citet{you81,pat84,haw90}    \\ 
             &            \citet{rin96,spr96,tap97}     \\         
             &            \citet{win03,god06}             \\ 
YY Dra       & see \citet{muk12} and references therein      \\ 
\enddata
\\ 
\small{
Note: Some systems (e.g. SS Cyg) have many more references 
and we list here only a few of the most relevant} 
\end{deluxetable}

\clearpage

\begin{figure}[h] 
\plotone{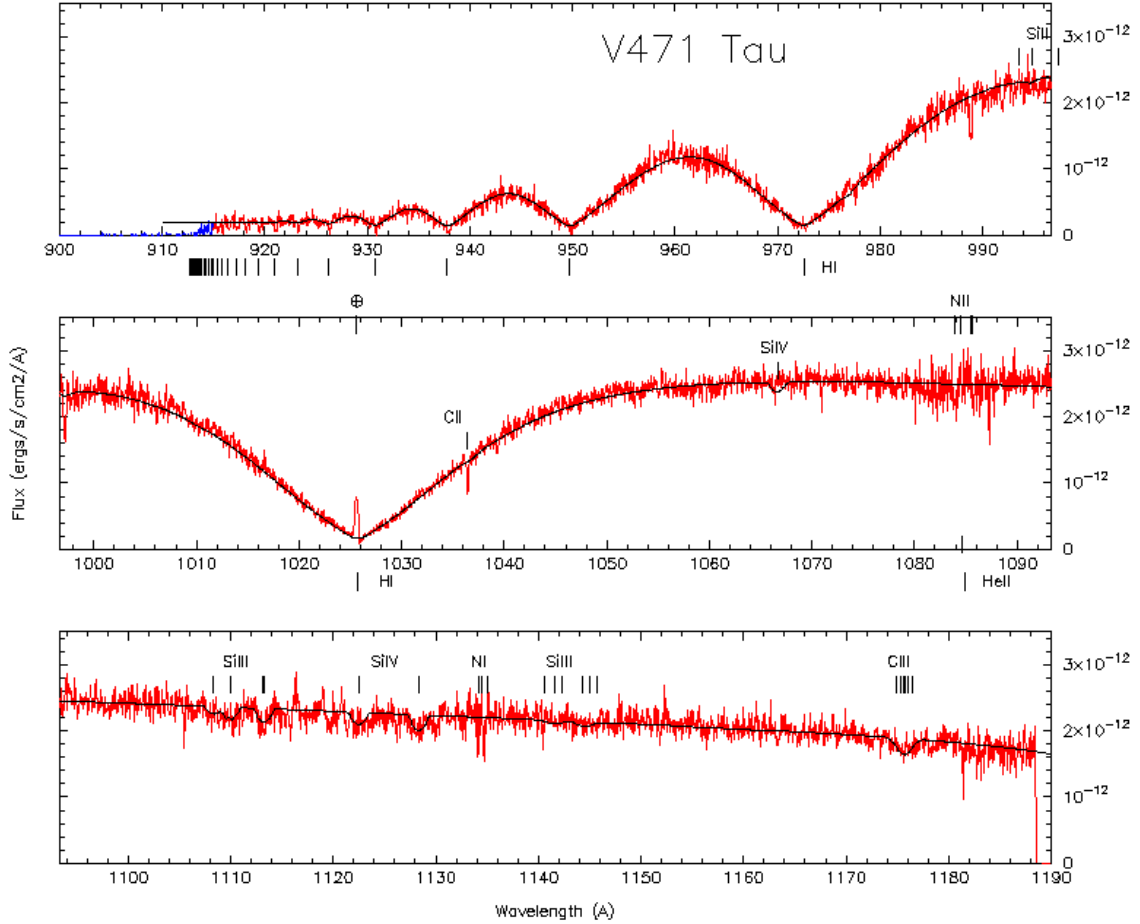}          
\caption{The {\it FUSE} spectrum of the WD in 
V471 Tau (in red) is fitted with 
a WD atmosphere model (in black). The WD atmosphere model is composed
mainly of hydrogen and the Lyman series (marked below each panel) 
is clearly seen. The WD model has a gravity $log(g)=8.3$, a temperature
of 33,400~K, a carbon abundances of $3 \times 10^{-4}$ solar, and a 
silicon abundance of $1.5 \times 10^{-2}$ solar.  The rotational velocity
used to match the shape of the lines is 250~km~$s^{-1}$ which is about a factor
of four larger than expected. It is possible that the broadening of the lines 
is due to a weak Zeeman effect.  
This is a good example of an almost
pure Hydrogen stellar atmosphere spectrum in which the main features
are due to the absorption lines of the Hydrogen Lyman series.  
} 
\end{figure} 

\clearpage 

\begin{figure}[h] 
\plotone{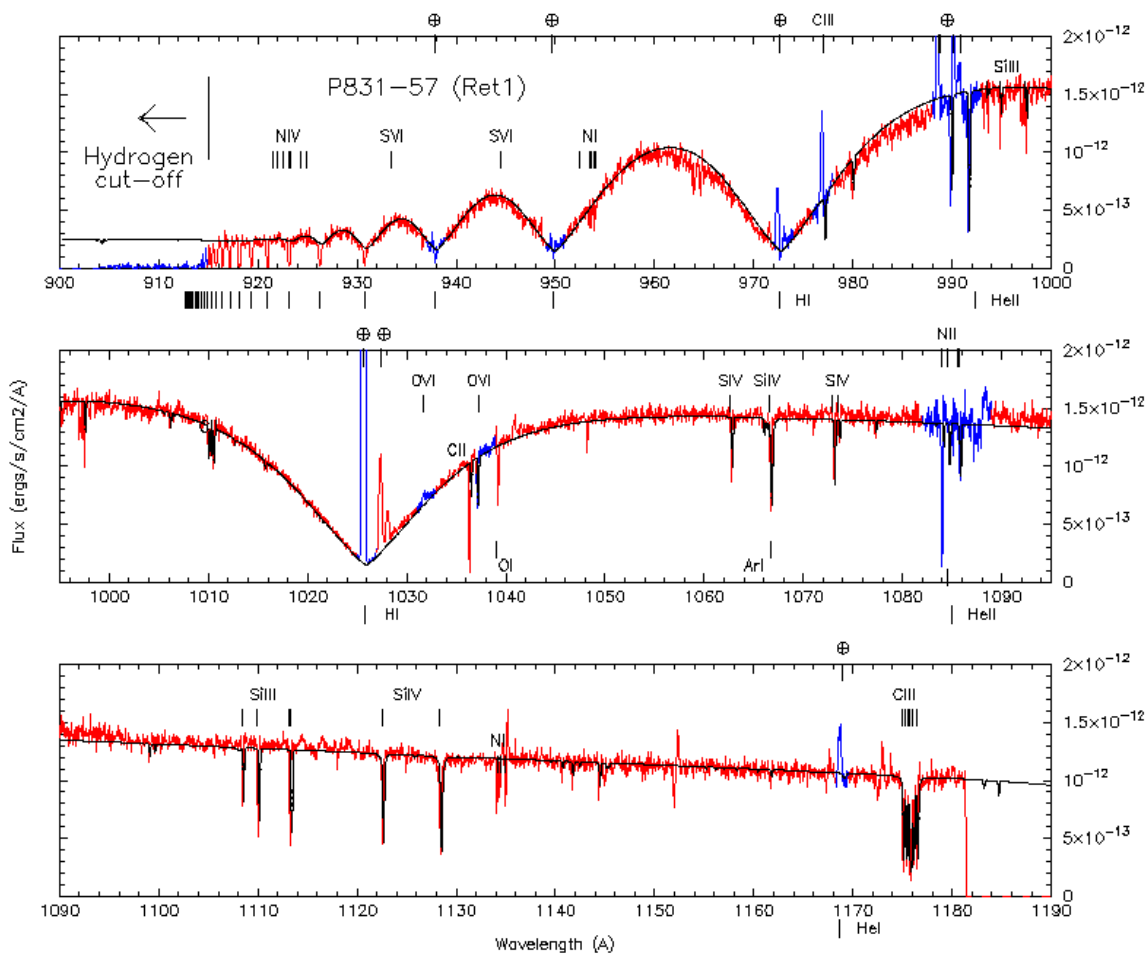}          
\caption{The {\it FUSE} spectrum of the wind accreting WD in P831-57 
(in red) is fitted with a WD atmosphere model (in black). Regions that
have been masked before the fitting are in blue. The WD model has 
a gravity of $log(g)=7.85$, a temperature of 37,500~K, carbon, sulfur,
and nitrogen abundances of 1\% solar, and silicon abundance of 2\% solar. 
The stellar rotational broadening has been set to only a few km~$s^{-1}$
in order to fit the sharp absorption lines. Airglow emissions are 
marked with a cross inside a circle. Compared to Fig.1, one can see
how a slight increase in chemical abundances affects a WD stellar spectrum.  
} 
\end{figure} 

\clearpage

\begin{figure}[h] 
\vspace{-5.cm} 
\plotone{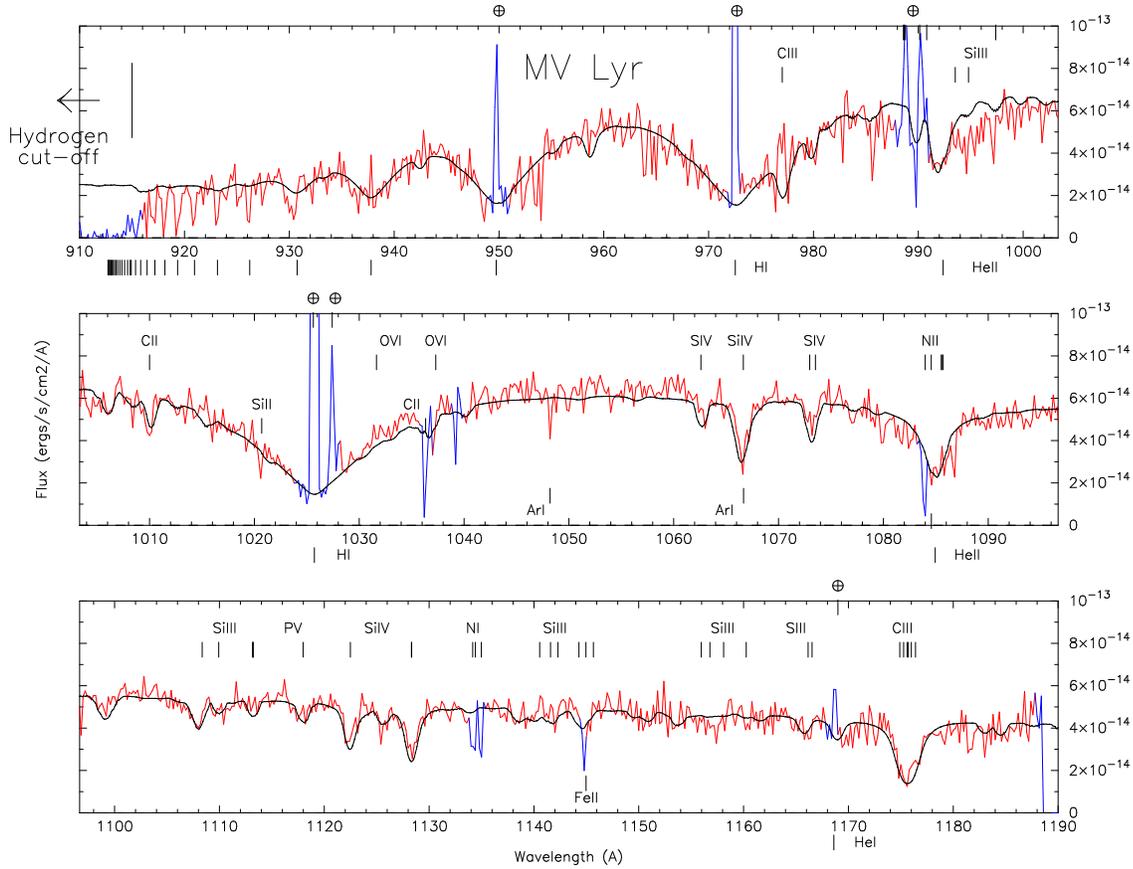}          
\caption{The {\it FUSE} spectrum of the Nova-like MV Lyr.     
This spectrum was obtained when MV Lyr dropped into a low state,
therefore revealing the WD. The WD theoretical model has a surface
gravity of $\log(g)=8.01$, a surface temperature $T_{eff}=45,000$~K,
solar composition, and a rotational velocity of 200~km~$s^{-1}$. The model 
fit gives a distance of 500~pc.  These results are in good agreement
with the modeling of \citet{hoa04}. This is a good example of a 
WD stellar atmosphere with solar composition and a moderate stellar
rotation where the (metal) absorption lines significantly affect
the continuum.   
} 
\end{figure}

\clearpage

\begin{figure}[h] 
\vspace{-2.cm} 
\plotone{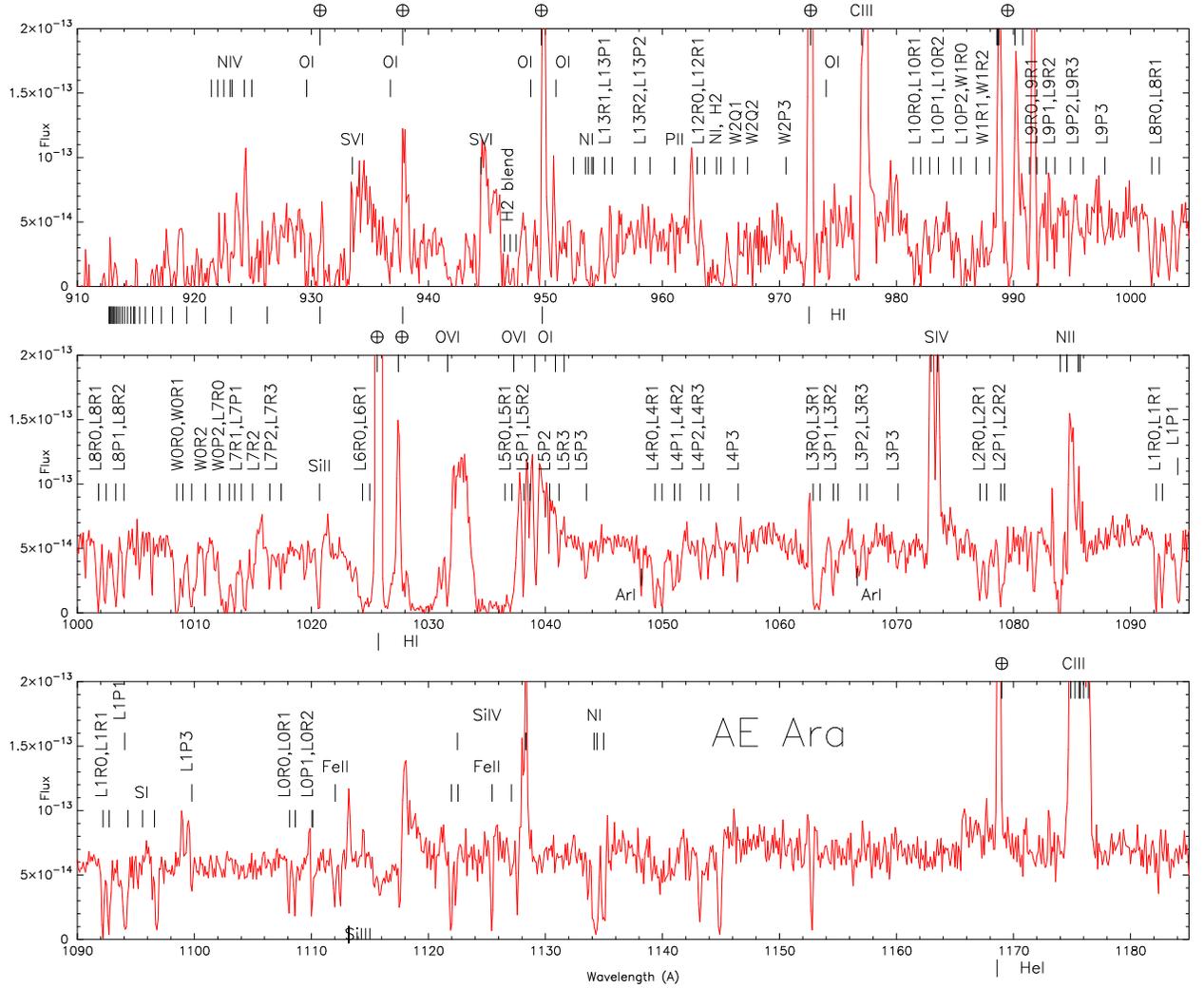}          
\caption{The {\it FUSE} spectrum of the symbiotic star AE Ara 
illustrates the multitude of ISM absorption
lines that basically slice the spectrum starting around 1100~\AA\  
all the way to the shortest wavelengths. The most prominent  
lines are identified by their band (Werner/W or Lyman/L), upper vibrational
level (1-13), and rotational transition ($R$, $P$, or $Q$ with lower rotational
state $J=1-3$). Additional ISM {\it metal} lines seen in the {\it FUSE} spectra
of our catalog include lines from Ar\,{\sc i}, 
Fe\,{\sc ii}, N\,{\sc i}, O\,{\sc i}, P\,{\sc ii}, Si\,{\sc ii}, S\,{\sc iii},
C\,{\sc ii}, many of which are not present in the above spectrum
of AE Ara.   
} 
\end{figure}

\clearpage

\begin{figure}[h]
\vspace{-2.cm} 
\plotone{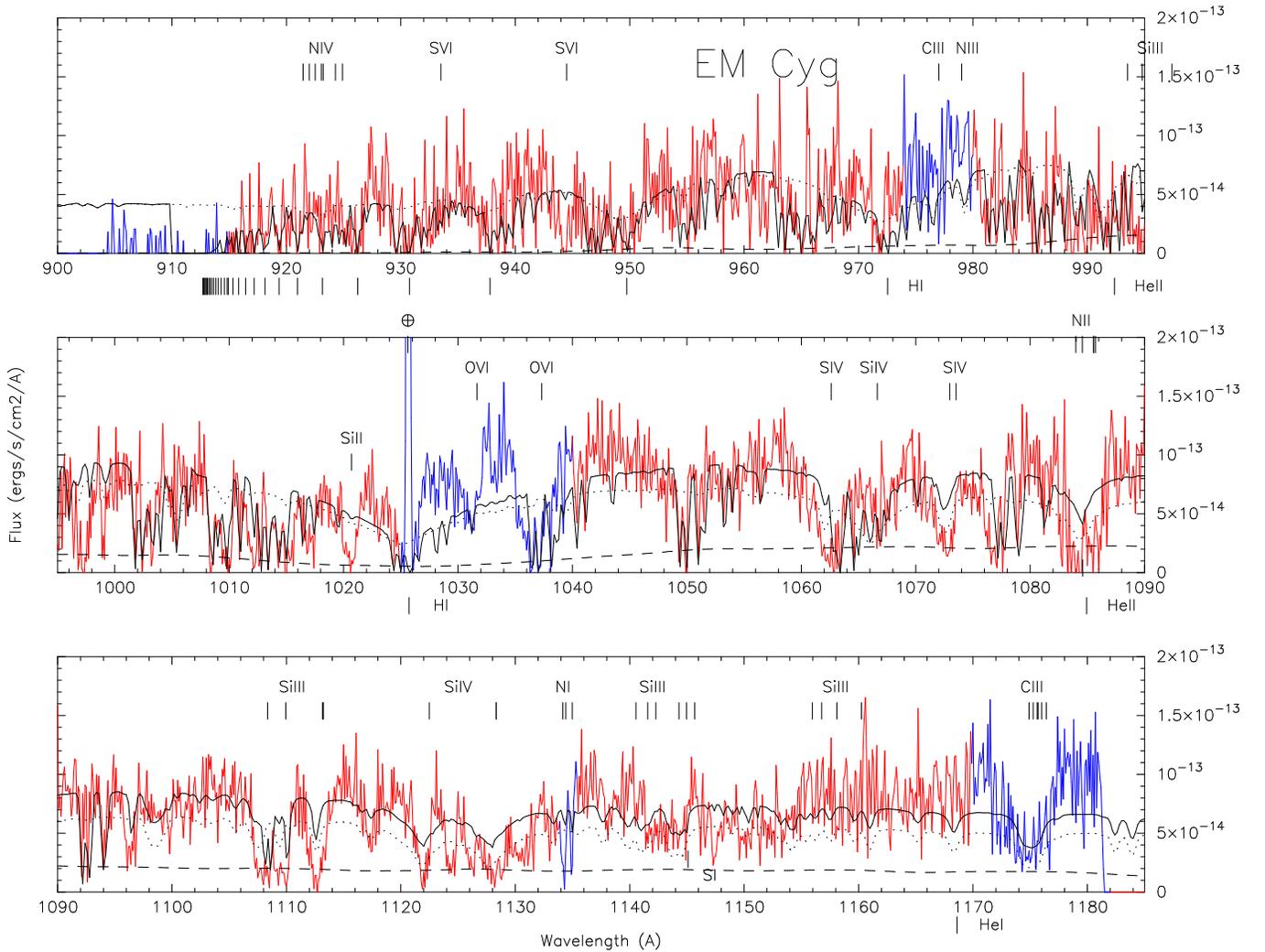}
\caption{The {\it FUSE} spectrum (exposure 1) of the EM Cyg is 
shown with a WD+disk model fit (solid black line) 
including ISM absorption modeling. The $1 M_{\odot}$ WD 
model has a temperature of 50,000~K, a projected rotational velocity of
100~km~$s^{-1}$, over-solar abundances of Si (30), S(10) and N(10).
The disk model has $\dot{M}=1 \times 10^{-10}~ M_{\odot}$/yr with
an inclination of $60^{\circ}$. The WD (dotted line) contributes 93\% of
the flux and the disk (dashed line) contributes the remaining  7\%.   
}
\end{figure}

\clearpage

\begin{figure}[h]
\vspace{-6.cm} 
\plotone{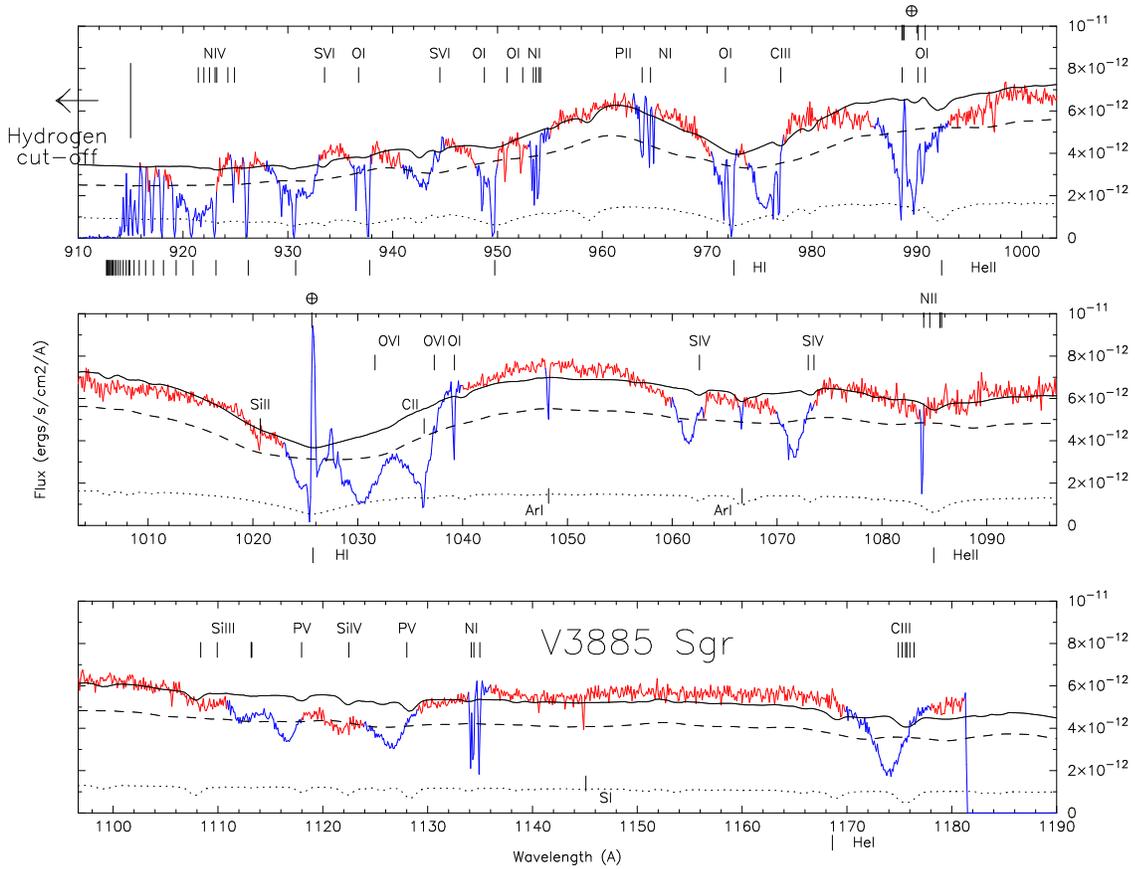}
\caption{The {\it FUSE} spectrum of the nova-like V3885 Sgr is 
shown with a WD+disk model fit (solid black line) 
including the contribution from a hot
boundary layer.  The $0.7M_{\odot}$ WD 
model has a temperature of 60,000~K, a projected rotational velocity of
200~km~$s^{-1}$, solar abundances.
The disk model has $\dot{M}=3 \times 10^{-9}~ M_{\odot}$/yr with
an inclination of $60^{\circ}$ and includes two inner (boundary layer) rings
with T=175,000~K. The WD (dotted line) contributes only 22\% of
the flux and the disk+BL (dashed line) contributes 78\%. 
}
\end{figure}


\begin{thebibliography} 

\bibitem[Abgrall et al. (2000)]{abg00} 
Abgrall, H., Roueff, E., \& Drira, I. 2000, \aaps, 141, 297. 

\bibitem[Allen (1978)]{all78}  
Allen, D.A. 1978, \mnras, 184, 601 

\bibitem[Allen (1980)]{all80}  
Allen, D.A. 1980, \mnras, 192, 521 

\bibitem[Allen \& Glass (1974)]{all74}  
Allen, D.A. \& Glass, I.S. 1974, \mnras, 167, 337  

\bibitem[Andronov (1991)]{and91}  
Andronov, I.L. 1991, Commission 27 of the IAU, Information Bulletin
on Variable Stars, p.3645  

\bibitem[Andronov \& Chinarova (1995)]{and95}  
Andronov, I.L., \& Chinarova, L.L. 1995 ASSL 205, 150  

\bibitem[Andronov \& Chinarova (1996)]{and96}  
Andronov, I.L., \& Chinarova, L.L. 1996, OAP 9, 9  

\bibitem[Andronov et al. (2003)]{and03}  
Andronov, I.L. et al. 2003, ASP Conference Series, Vol. 292, Interplay
of Periodic, Cyclic and Stochastic Variability in Selected Areas of the 
H-R Diagram, ed. C. Sterken, p.313  

\bibitem[Barrett et al. (2012)]{bar12}  
Barrett, P.E., Dupuis, J., Schlegel, E.M., Heathcote, B.D., Godon, P., 
Sion, E.M. 2012, ApJ, submitted 

\bibitem[Barstow et al. (1992)]{bar92}  
Barstow, M.A., Schmitt, J.H.M.M., Clemens, J.C., Pye, J.P.,
Denby, M., Harris, A.W., Pankiewicz, G.S. 1992, \mnras, 255, 369  

\bibitem[Barstow et al. (1997)]{bar97}  
Barstow, M.A., Holberg, J.B., Cruise, A.M., Penny, A.J., 
1997, \mnras, 290, 505 

\bibitem[Bateson (1979)]{bat79}  
Bateson, F.M., 1979, Ro.Astron.Soc. of New Zealand - 
Publ. Variable Star Sect., 7, 20

\bibitem[Bateson \& McIntosh (1990)]{bat90}  
Bateson, F.M., \& McIntosh, R. 1990, Ro.Astron.Soc. of New Zealand - 
Publ. of Variables Star Section, 17, 76  

\bibitem[Bayley (1981)]{bay81}  
Bayley, J.A. 1981, \mnras, 197, 31 

\bibitem[Becker et al. (1982)]{bec82}  
Becker, R.H., Chanan, G.A., Wilson, A.S., Pravdo, S.H. 1982, MNRAS, 201, 265 

\bibitem[Belczy\'nski et al. (2000)]{bel00} 
Belczy\'nski, K., Miko\/lajewska, J., 
Brandi, E., Garcia, L.G., Ferrer, O.E., Pereira, C.B.  
2000, New AR, 44, 81 

\bibitem[Belloni et al. (1991)]{bel91}  
Belloni, T. et al. 1991, \aap, 246, L44 

\bibitem[Bengtsson (1975)]{ben75}  
Bengtsson, H., 1975, Astronomisk Tidsskrift 8, 74 

\bibitem[Beuermann \& Pakull (1984)]{beu84}  
Beuermann, K. \& Pakull, M.W. 1984, \aap, 136, 250 

\bibitem[Beuermann \& Thomas (1990)]{beu90}  
Beuermann, K., \& Thomas, H.-C. 1990, \aap, 230, 326  

\bibitem[Beuermann et al. (1992)]{beu92} 
Beuermann, K., Stasiewski, U., \& Schwope, A.D. 1992, \aap, 256, 433  

\bibitem[Beyer (1967)]{bey67} 
Beyer, M., 1967, Astronomische Nachrichten, 290, 29 

\bibitem[Bitner et al. (2007)]{bit07}  
Bitner, M.A., Robinson, E.L., \& Behr, B.B. 2007, \apj, 662, 564  

\bibitem[Bochkarev \& Sitnik (1985)]{boc85}  
Bochkarev, N.G., \&  Sitnik, T.G. 1985, \apss, 108, 237  

\bibitem[Bois et al. (1988)]{boi88}  
Bois, B., Mochnacki, S.W., Lanning, H.H. 1988, \aj, 96, 157 

\bibitem[Bolick et al. (1987)]{bol87} 
Bolick, U., Beuermann, K., Bruch, A., Lenzen, R. 1987, \apss, 130, 175 

\bibitem[Bond et al. (2001)]{bon01}  
Bond, H.E., Mullan, D., O'Brien, S., \& Sion, E.M. 2001, \apj, 560, 919  

\bibitem[Brady \& Herczeg (1977)]{bra77}  
Brady, R.A., Herczeg, T.G. 1977, \pasp, 89, 71 

\bibitem[Bruch (1989)]{bru89}  
Bruch, A. 1989, \aaps, 78, 145  

\bibitem[Bruch \& Engel (1994)]{bru94}
Bruch, A., \& Engel, A. 1994, \aaps, 104, 79  

\bibitem[Bruch et al. (1987)]{bru87}  
Bruch, A., Fischer, F.-J., Wilmsen, U. 1987, \aap, 70, 481  

\bibitem[Burbidge \& Burbidge (1953)]{bur53}  
Burbidge, E.M., Burgidge, G.R. 1953, \apj, 118, 349 

\bibitem[Cannizzo (1992)]{can92}  
Cannizzo, J.K., \& Mattei, J.A. 1992, \apj, 401, 642 

\bibitem[Cannizzo (1993)]{can93}  
Cannizzo, J.K. 1993, \apj, 419, 318  

\bibitem[Cannon (1933)]{can33}  
Cannon, A.J. 1933, Harvard Obs. Bull., 891, 9  

\bibitem[Cester \& Pucillo (1976)]{ces76}  
Cester, B. \&  Pucillo, M. 1976, \aap, 46, 197 

\bibitem[Cheng et al. (2001)]{che01}  
Cheng, A., O'Donoghue, D., Stobie, R.S., Kilkenny, D., Warner, B., 2001,
\mnras, 325, 89  

\bibitem[Chiappetti et al. (1982)]{sch82}  
Chiappetti, L., Maraschi, L., Treves, A., Tanzi, E.G. 1982, \apj, 258, 236  

\bibitem[Clemens et al. (1992)]{cle92}  
Clemens, J.C. et al. 1992, \apj, 391, 773  

\bibitem[Costero et al. (1998)]{cos98}  
Costero, R., Echevarria, J., \& Pineda, L. 1998, AAS, 192, 8206

\bibitem[Costero et al. (2004)]{cos04}  
Costero, R., Echevarria, J., Michel, R., Zharikov, S. 2004, AAS, 205, 1906 

\bibitem[Csizmadia et al. (2008)]{csi08}  
Csizmadia, Sz., Nagy, Sz., Borkovits, T., Heged\"us, T., Bir\'o, I.B.,
Kiss, Z.T. 2008, Astronomische Nachrichten, 329, 39  

\bibitem[Davis \& Mattei (1981)]{dav81}  
Davis, J.F., Mattei, J.A. 1981, AAVSO Journal, 10, 28  

\bibitem[Dixon et al. (2007)]{dix07} 
Dixon, W.V., et al. 2007, \pasp, 119, 527  

\bibitem[Downes et al. (2001)]{dow01}
Downes, R.A., Webbink, R.F., Shara, M.M., Ritter, H., Kolb, U.,
D\"urbeck, H.W. 2001, \pasp, 113, 764 

\bibitem[Dupuis et al. (1997)]{dup97}  
Dupuis, J., Vennes, S., Chayer, P., Cully, S., \& Rodriguez-Bell, T. 
1997, in White Dwarfs, Eds J. Isern, M. Hernanz \& E. Garcia-Berro (Dordrecht:
Kluwer), 375  

\bibitem[Eggen \& Niemela (1984)]{egg84}  
Eggen, O.J., \&  Niemela, V.S. 1984, \aj, 89, 389  

\bibitem[Fekel et al. (2003)]{fek03} 
Fekel, F.C., Hinkle, K.H., Joyce, R.R. 2003, ASPC 303, 113  

\bibitem[Fekel et al. (2010)]{fek10}  
Fekel, F.C., Hinkle, K.H., Joyce, R.R., Wood, P.R., 2010, \aj, 139, 1315 

\bibitem[Friend et al. (1990)]{fri90}  
Friend, M.T., Martin, J.S., Connon-Smith, R., Jones, D.H.P. 1990, 
\mnras, 246, 654 

\bibitem[Froning et al. (2012)]{fro12}
Froning, C.S., Long, K.S., G\"ansicke, B.T., Szkody, P. \apjs, 199, 7  

\bibitem[Galis et al. (1999)]{gal99}  
Galis, R., Hric, L., Friedjung, M., Petrik, K. 1999, \aap, 348, 533 

\bibitem[G\"ansicke \& Beuermann (1996)]{gan96}  
G\"ansicke, B.T., \&  Beuermann, K. 1996, \aap, 309, L47 

\bibitem[G\"ansicke et al. (1997)]{gan97}  
G\"ansicke, B.T., Beuermann, K., \& Thomas, H.-C. 1997, \mnras, 289, 388 

\bibitem[G\"ansicke et al. (2006)]{gan06} 
G\"ansicke, B.T., Long, K.S., Barstow, M.A., Hubeny, I. 2006, \apj, 639, 1039 

\bibitem[G\"ansicke et al. (2003)]{gan03}  
G\"ansicke, B.T., Szkody, P., de Martino, D., Beuermann, K., 
Long, K.S., Sion, E.M., Knigge, C., Marsh, T., \& Hubeny, I. 2003, 
\apj, 594, 443 

\bibitem[G\"ansicke et al. (2001)]{gan01}  
G\"ansicke, B.T., Szkody, P., Sion, E.M., Hoard, D.W., Howell, S., 
Cheng, F.H., \& Hubeny, I. 2001, \apj, 374, 656 

\bibitem[Garrison et al. (1982)]{gar82}  
Garrison, R.F., Hiltner, W.A., \& Schild, R.E. 1982, IAU Circ. 3730, 2   

\bibitem[Garrison et al. (1984)]{gar84}  
Garrison, R.F., Schild, R.E., Hiltner, W.A., Krzeminski, W. 1984, 
\apj, 276, L13  

\bibitem[Gessner (1978)]{ges78}  
Gessner, H. 1978, Mitt. Veraenderl. Sterne, 8, 66  

\bibitem[Gilliland (1982)]{gil82}    
Gilliland, R. L. 1982, \apj, 263, 302 

\bibitem[Godon et al. (2006)]{god06} 
Godon, P., Seward, L., Sion, E.M., Szkody, P., 2006, \aj, 131, 2634 

\bibitem[Godon \& Sion (2005)]{god05}  
Godon, P. \&  Sion, E.M. 2005, \mnras, 361, 809 

\bibitem[Godon \& Sion (2011)]{god11}  
Godon, P., \& Sion, E.M. 2011, \pasp, 2011, 123, 903     

\bibitem[Godon et al. (2007)]{god07} 
Godon, P., Sion, E.M., Barrett, P.E., Szkody, P. 2007, \apj, 656, 1092 

\bibitem[Godon et al. (2008a)]{god08a} 
Godon, P., Sion, E.M., Barrett, P.E., Szkody, P., Schlegel, E.M. 2008a, \apj, 687, 532  

\bibitem[Godon et al. (2008b)]{god08b} 
Godon, P., Sion, E.M., Barrett, P.E., Hubeny, I., Linnell, A.P., Szkody, P. 2008b, \apj, 679, 1447 
 
\bibitem[Godon et al. (2009a)]{god09a} 
Godon, P., Sion, E.M., Barrett, P.E., Szkody, P., 2009a, \apj, 701, 1091

\bibitem[Godon et al. (2009b)]{god09b} 
Godon, P., Sion, E.M., Barrett, P.E., Linnell, A.P., 2009b, \apj, 699, 1229 

\bibitem[Godon et al. (2004)]{god04}  
Godon, P., Sion, E.M., Cheng, F.H., Szkody, P., Long, K.S., \& 
Froning, C.S. 2004, \apj, 612, 429

\bibitem[Gonz\'alez-Riestra et al. (1999)]{gon99}  
Gonz\'alez-Riestra, R., Viotti, R., Iijima, T., Greiner, J. 1999, \aap, 347, 478  

\bibitem[Green et al. (1986)]{gre86}  
Green, R.F., Schmidt, M., \& Liebert, J. 1986, \apjs, 61, 305  

\bibitem[Greenstein \& Oke (1982)]{gre82} 
Greenstein, J.L., \& Oke, J.B. 1982, \apj, 258, 209  

\bibitem[Greiner et al. (1997)]{gre97}  
Greiner, J., Bickert, K., Luthardt, R., Viotti, R., Altamore, A., 
Gonz\'alez-Riestra, R., Stencel, R.E. 1997, \aap, 322, 576 

\bibitem[Guinan \& Sion (1982)]{gui82}  
Guinan, E.F., \& Sion, E.M. 1982, \apj, 258, 217  

\bibitem[Guinan \& Sion (1984)]{gui84}  
Guinan, E.F., \&  Sion, E.M. 1984, \aj, 89, 1252  

\bibitem[Hacke (1987)]{hac87}  
Hacke, G. 1987, Astronomicheskii Tsirkulyar, 1491, 6 
 
\bibitem[Hamada \& Salpeter (1961)]{ham61} 
Hamada, T., \& Salpeter, E.E. 1961, \apj, 134, 683; 

\bibitem[Hamilton et al. (2007)]{ham07}  
Hamilton, R.T., Urban, J.A., Sion, E.M., Riedel, A.R., Voyer, E.N., 
Marcy, J.T., Lakatos, S.L.  2007, \apj, 667, 1139   

\bibitem[Harrison et al. (1999)]{har99}  
Harrison, T.E., McNamara, B.J., Szkody, P., McArthur, B.E., 
Benedict, G.F., Klemola, A.R., Gilliland, R.L. 1999, \apj, 515, L93 

\bibitem[Harrison et al. (2000)]{har00}  
Harrison, T.E., McNamara, B.J., Szkody, P., \& Gililand, R.  2000,
\aj, 120, 2649  

\bibitem[Harrison et al. (2004)]{har04} 
Harrison, T.E., Johnson, J.J., McArthur, B.E., Benedict, G.F., Szkody, P.,
Howell, S.B., Gelino, D.M. 2004, \aj, 127 460 

\bibitem[Harrison et al. (2005)]{har05}  
Harrison, T.E., Osborne, H.L., Howell, S.B. 2005, \aj, 129, 2400 

\bibitem[Hartley et al. (2002)]{har02}  
Hartley, L.E., Drew, J.E., Long, K.S., Knigge, C., Proga, D.
2002, \mnras, 332, 127  

\bibitem[Hartley et al. (2005)]{hart05}  
Hartley, L.E., Murray, J.R., Drew, J.E., \& Long, K.S. 2005, \mnras, 363, 285  

\bibitem[Hassal (1985)]{has85}  
Hassal, B.J.M. 1985, \mnras, 216, 335  

\bibitem[Haug (1987)]{hau87}  
Haug, K. 1987, \apss, 130, 91  

\bibitem[Haug (1988)]{hau88}  
Haug, K. 1988, \mnras, 235, 1385  

\bibitem[Haug \& Drechsel (1985a)]{hau85}
Haug, K., \& Drechsel, H., 1985a, Mitteilungen der 
Astronomischen Gesellschaft, 63, 193 

\bibitem[Haug \& Drechsel (1985b)]{hau85b}  
Haug, K., \& Drechsel, H. 1985b, \aap, 151, 157 

\bibitem[Hawkins et al. (1990)]{haw90}  
Hawkins, M.R., Smith, R., \& Jones, D.H.P. 1990, in Accretion-Powered 
Compact Binaries, ed. C. Mauche (Cambridge: Cambridge Univ. Press), 113 

\bibitem[Herczeg (1987)]{her87}  
Herczeg, T.J. 1987, \apss, 130, 39 

\bibitem[Hoard et al. (2004)]{hoa04}  
Hoard, D.W., Linnell, A.P., Szkody, P., Fried, R.E., Sion, E.M., 
Hubeny, I., Wolfe, M.A., 2004, \apj, 604, 346 
  
\bibitem[Hoard et al. (2003)]{hoa03} 
Hoard, D.W., Szkody, P., Froning, C.S., Long, K.S., \& Knigge, C. 2003,
\aj, 126, 2473 
 
\bibitem[Hoffmeister (1928)]{hof28} 
Hoffmeister, C. 1928, Astron.Nach., 234, 33  

\bibitem[Holm (1988)]{hol88b}  
Holm, A.V. 1988, in ESA, A Decade of UVAstronomy with the IUE Satellite, 
ed. E.J.Rolf (ESA SP-281; Paris: ESA), 229  

\bibitem[Holm \& Polidan (1988)]{hol88a}  
Holm, A.V., \& Polidan, R.S. 1988, in ESA, A Decade of UVAstronomy 
with the IUE Satellite, ed. E.J.Rolf (ESA SP-281; Paris: ESA), 179  

\bibitem[Honeycutt \& Schlegel (1985)]{hon85}  
Honeycutt, R.K., \&  Schlegel, E.M. 1985, \pasp, 97, 1189  

\bibitem[Honeycutt et al. (1994)]{hon94}  
Honeycutt, R.K., Cannizzo, J.K., \&  Robertson, J.W. 1994, \apj, 425, 835

\bibitem[Honeycutt \& Robertson (1998)]{hon98}   
Honeycutt, R.K., \&  Robertson, J.W. 1998, \apj, 116, 1961  

\bibitem[Howarth (1976)]{how76}  
Howarth, I.D. 1976, Mitt. Veraenderl. Sterne, 7, 147 

\bibitem[Howell \& Szkody (1990)]{how90}  
Howell, S.B., \& Szkody P. 1990, \apj, 356, 623  

\bibitem[Huang et al. (1996a)]{hua96a}  
Huang, M., Sion, E.M., Hubeny, I., Cheng, F., \& Szkody, P. 1996a, 
\apj, 458, 355

\bibitem[Huang et al. (1996b)]{hua96b}  
Huang, M., Sion, E.M., Hubeny, I., Cheng, F.-H., \& Szkody, P. 1996b, 
\aj, 111, 2386

\bibitem[Hubeny (1988)]{hub88} 
Hubeny, I. 1988, Comput. Phys. Commun., 52, 103; 

\bibitem[Hubeny \& Lanz (1995)]{hub95} 
Hubeny, I.,\& Lanz, T. 1995, \apj, 439, 875 

\bibitem[\.{I}bano\u{g}lu (1978)]{iba78}                 
\.{I}bano\u{g}lu, C. 1978, \apss, 57, 219 

\bibitem[\.{I}bano\u{g}lu et al. (2005)]{iba05}  
\.{I}bano\u{g}lu, C., Evren, S., Ta\c{s}, G., \c{C}ak{\i}rl{\i}, \"O. 
2005, \mnras, 360, 1077  

\bibitem[Isles (1976)]{isl76}  
Isles, J.E., 1976, Journal of the British Astronomical Association, 86, 412 

\bibitem[Jameson et al. (1981)]{jam81}  
Jameson, R.F., King, A.R., Sherrington, M.R. 1981, \mnras, 195, 235 

\bibitem[Jeffries et al. (2010)]{jef10}  
Jeffries, R., Jackson, K., Briggs, P., Evans, J.P. 2010, \mnras, 411, 2029  

\bibitem[Johnson et al. (2003)]{joh03}  
Johnson, J.J., Harrison, T.E., Howell, S.B., Szkody, P., McArthur, B.E., 
Benedict, G.F. 2003, BAAS, 202, 0702

\bibitem[Kato et al. (2002)]{kat02} 
Kato, T. et al. 2002, \aap, 396, 929  

\bibitem[Kato et al. (2004)]{kat04} 
Kato, T. et al. 2004, \mnras, 347, 861    

\bibitem[Knigge (2006)]{kni06} 
Knigge, C. 2006, \mnras, 373, 484 

\bibitem[Knigge (2007)]{kni07} 
Knigge, C. 2007, \mnras, 382, 1982  

\bibitem[Kraft (1964)]{kra64} 
Kraft, R.P. 1964, 1st Conf. on Faint Blue Stars, ed. Luyten, U. of 
Minnesota Press, Minneapolis 

\bibitem[Kubiak et al. (1999)]{kub99}  
Kubiak, M., Pojmanski, G., Krzeminski, W. 1999, Acta Astronomica, 49, 73  

\bibitem[La Dous (1991)]{lad91}
La Dous, C. 1991, \aap, 252, 100 

\bibitem[Lasota (2001)]{las01} 
Lasota, J.-P. 2001, New Astron. Rev., 45, 449  

\bibitem[Lake \& Sion (2001)]{lak01}  
Lake, J., \&  Sion, E.M. 2001, \aj, 122, 1632  

\bibitem[Lesniak \& Sion (2003)]{les03}  
Lesniak, M.V., \& Sion, E.M. 2003, AAS, 203, 4412  

\bibitem[Linnell et al. (2005)]{lin05}  
Linnell, A.P., Szkody, P., G\"ansicke, B.T., Long, K.S., Sion, E.M., 
Hoard, D.W., Hubeny, I.  2005, \apj, 624, 923 

\bibitem[Linnell et al. (2007)]{lin07} 
Linnell, A.P., Godon, P., Hubeny, I., Sion, E.M., Szkody, P. 2007, \apj, 662, 1204  

\bibitem[Linnell et al. (2008a)]{lin08a} 
Linnell, A.P., Godon, P., Hubeny, I., Sion, E.M., Szkody, P. 
\& Barrett, P.E. 2008a, \apj, 676, 1226  

\bibitem[Linnell et al. (2008b)]{lin08b} 
Linnell, A.P., Godon, P., Hubeny, I., Sion, E.M., Szkody, P. 2008b, 
\apj, 688, 568  

\bibitem[Linnell et al. (2009)]{lin09} 
Linnell, A.P., Godon, P., Hubeny, I., Sion, E.M., Szkody, P. 
\& Barrett, P.E. 2009, \apj, 703, 1839  

\bibitem[Linnell et al. (2010)]{lin10} 
Linnell, A.P., Godon, P., Hubeny, I., Sion, E.M., Szkody, P. 2010, 
\apj, 719, 271  

\bibitem[Long et al. (1996)]{lon96}  
Long, K.S., Blair, W., Hubeny, I., \& Raymond, J. 1996, \apj, 468, 871 

\bibitem[Long et al. (2007)]{lon07}   
Long, K.S., Froning, C.S., G\"ansicke, B.T., \& Knigge, C. 2007, ASPC, 372, 505  

\bibitem[Long et al. (2005)]{lon05}  
Long, K.S., Froning, C.S., Knigge, C., Blair, W.P., Kallman, T.R., 
Ko, Y.-K. 2005, \apj, 630, 511  
\bibitem[Long et al. (2009)]{lon09}
Long, K.S., G\"ansicke, B.T., Knigge, C., Froning, C.S., Monard, B. 2009,
\apj, 697, 1512 

\bibitem[Long et al. (1994)]{lon94}  
Long, K.S., Wade, R.A., Blair, W.P., Davidsen, A.F., Hubeny, I.
1994, \apj, 426, 704   

\bibitem[Mateo \& Szkody (1984)]{mat84}  
Mateo, M., \&  Szkody, P. 1984, \aj, 89, 863 

\bibitem[Mauche (1991)]{mau91}  
Mauche, C.W. 1991, \apj, 373, 624 

\bibitem[Mauche (1996)]{mau96}  
Mauche, C.W. 1996, in A. Evans and J.H. Woods (eds.), Cataclysmic 
Variables and Related Objects, 243 (Kluwer Academic Publishers, 
The Netherlands)

\bibitem[Mauche et al. (1988)]{mau88} 
Mauche, C.W., Raymond, J.C., \& C\'ordova, F.A. 1988, \apj, 335, 829  

\bibitem[Mauche et al. (1991)]{mau91e}  
Mauche, C., Wade, R.A, Polidan, R.S., van der Woerd, H., \& 
Paerels, F.B.S. 1991, \apj, 372, 659

\bibitem[Meinunger (1979)]{mei79}  
Meinunger, L. 1979, Inf.Bull.Variable Stars, 1611, 1   

\bibitem[Menzies et al. (1986)]{men86}  
Menzies, J.W., Odonoghue, D., Warner, B. 1986, \apss, 122, 73 

\bibitem[Metz (1989)]{met89}  
Metz, K. 1989, Inf. Bull. Var. Stars, 3385  

\bibitem[Miko\/lajewska et al. (1997)]{mik97}  
Miko\/lajewska, J., Acker, A., \& Stenholm, B. 1997, \aap, 327, 191  

\bibitem[Miko\/lajewska et al. (1999)]{mik99}  
Miko\/lajewska, J., Belczy\'nski, K., Brandi, E., Garcia, L.G., Ferrer, O.E. 
1999, BAAA, 43, 31 

\bibitem[Miko\/lajewska et al. (1989)]{mik89}  
Miko\/lajewska, J., Kenyon, S.J., \& Miko\/lajewski, M. 1989, \aj, 98, 1427  

\bibitem[Mikolajewska et al. (1995)]{mik95}  
Miko\/lajewska, J., Kenyon, S.J., Miko\/lajewski, M., Garcia, M.R., Polidan, R.S, 
1995, \aj, 109, 1289  

\bibitem[Miko\/lajewska et al. (2003)]{mik03}  
Miko\/lajewska, J., Quiroga, C., Brandi, E., Garcia, L.G., Ferrer, O.E., 
\& Belczy\'nski, K., 2003, in Symbiotic Stars, Probing Stellar Evolution,
ed. R.L.M. Corradi, J. Miko\/lajewska, \& T.J. Mahoney (San Francisco, CA: 
ASP) 147 

\bibitem[Moos et al. (2000)]{moo00}
Moos, H.W., et al. 2000, \apj, 538, L1 

\bibitem[Morton (2000)]{mor00} 
Morton, D.C. 2000, \apjs, 130, 403  

\bibitem[Morton (2003)]{mor03} 
Morton, D.C. 2003, \apjs, 149, 205, 

\bibitem[Mukai (2012)]{muk12}
Mukai, K. 2012, online catalog of IPs 
asd.gsfc.nasa.gov/Koji.Mukai/iphome/iphome.html 

\bibitem[Mumford (1979)]{mum79}  
Mumford, G.S. 1979, BAAS 11, 403 

\bibitem[Mumford (1980)]{mum80}  
Mumford, G.S 1980, \aj, 85, 748 

\bibitem[Mumford \& Krezminski (1969)]{mum69} 
Mumford, G.S., \& Krezminski, W. 1969, \apjs, 18, 429 

\bibitem[M\"urset et al. (1991)]{mur91}  
M\"urset, U., Nussbaumer, H., Schmid, H.M., \& Vogel, M. 1991, \aap, 248, 458  

\bibitem[North et al. (2000a)]{nor00a}  
North, R.C., Marsh, T.R., Kolb, U., Stehle, R., Smith, R.C., 
2000a, NewAR 44, 29 

\bibitem[North et al. (2000b)]{nor00b}  
North, R.C., et al. 2000b, \mnras, 313, 383 

\bibitem[North et al. (2002)]{nor02}  
North, R.C., Marsh, R.R., Kolb, U., Dhillon, V.S., Moran, C.K.J.
2002, \mnras, 337, 1215 

\bibitem[Notni \& Richter (1984)]{not84} 
Notni, P., \&  Richter, G.A. 1984, Inf.Bull.Variable Stars No.2634  

\bibitem[O'Brien et al. (2001)]{obr01}  
O'Brien, S., Bond, H.E., Sion, E.M. 2001, \apj, 563, 971  

\bibitem[O'Connell (1932)]{oco32}  
O'Connell, D.J.K. 1932, Harvard College Observatory Bulletin, 890, 180  

\bibitem[Pandel et al. (2003)]{pan03}  
Pandel, D., Cordova, F.A., Howell, S.B. 2003, \mnras, 346, 1231

\bibitem[Panei et al. (2000)]{pan00} 
Panei, J.A., Althaus, L.G., \& Benvenuto, O.G. 2000, \aap, 353, 970 

\bibitem[Patterson (1984)]{pat84} 
Patterson, J. 1984, \apjs, 54, 443  

\bibitem[Patterson (2002)]{pat02}  
Patterson, J. 2002, CBAstro Communication, (March 28;  

http://cbastro.org/communications/news/messages/0249.html)            

\bibitem[Patterson et al. (2005a)]{pat05a}  
Patterson, J. et al. 2005, \pasp, 117, 1204                                

\bibitem[Patterson et al. (2005b)]{pat05b}  
Patterson, J., Thorstensen, J.R., \& Kemp, J. 2005, \pasp, 117, 427

\bibitem[Perryman et al. (1997)]{per97} 
Perryman, M.A.C., et al. 1997, \aap, 323L, 49  

\bibitem[Polidan \& Carone (1986)]{pol86}
Polidan, R.S., \& Carone, T.E., 1986, BAAS 18, 1016 

\bibitem[Polidan et al. (1990)]{pol90}  
Polidan, R.S., Mauche, C.W., \& Wade, R.A. 1990, \apj, 356, 211  

\bibitem[Press et al. (1992)]{pre92} 
Press, W.H., Teukolsky, S.A., Vetterling, W.T., Flannery, B.P.,
Numerical Recipes in Fortran 77, The Art of Scientific Computing,
Second Edition, 1992, Cambridge University Press

\bibitem[Pretorius et al. (2006)]{pre06}  
Pretorius, M.L., Warner, B., \&  Woudt, P.A. 2006, \mnras, 368, 361 

\bibitem[Pringle (1975)]{pri75}  
Pringle, J. 1975, \mnras, 170, 633 

\bibitem[Prinja \& Rosen (1995)]{pri95}  
Prinja, R.K., \&  Rosen, R. 1995, \mnras, 273, 461  

\bibitem[Prinja et al. (2003)]{pri03} 
Prinja, R.K., Long, K.S., Froning, C.S., Knigge, C., Witherick, D.K., 
Clark, J.S., \& Ringwald, F.A. 2003, \mnras, 340, 551  

\bibitem[Provencal et al. (1996)]{pro96}  
Provencal, J., Shipman, H.L., Hoeg, E., Thejll, P. 1996, AAS 189.7901 

\bibitem[Puebla et al. (2007)]{pue07}
Puebla, R.E., Diaz, M.P., Hubeny, I. 2007, \aj, 134, 1923 

\bibitem[Ramseyer et al. (1995)]{ram95}  
Ramseyer, T.F., Hatzes, A.P., \& Jablonski, F. 1995, \aj, 110, 1364  

\bibitem[Ribeiro \& Diaz (2007)]{rib07}  
Ribeiro, F.M.A., \& Diaz, M.P. 2007, \aj, 133, 2659  

\bibitem[Richter (1961)]{ric61} 
Richter, G.A. 1961, Veroeff.Sternw.Sonneberg, 4, 433  

\bibitem[Richter et al. (1988)]{ric88} 
Richter, G.A., Notni, P., Tiersch, H. 1988, Astron.Nach., 309, 91  

\bibitem[Ringwald (1994)]{rin94}  
Ringwald, R.A. 1994, Interacting Binary Stars, ASP Conferences Series,
Vol.56, A.W. Shafter (ed.), p.294   

\bibitem[Ringwald et al. (1996)]{rin96}  
Ringwald, F.A., Thorstensen, J., Honeycutt, R., \& Smith, R.C. 
1996, \aj, 111, 2077

\bibitem[Ritter \& Kolb (2003)]{rit03}
Ritter, H., \& Kolb, U. 2003, \aap, 404, 301  

\bibitem[Robinson (1974)]{rob74}  
Robinson, E.L. 1974, \apj, 193, 191  

\bibitem[Robinson et al. (1981)]{rob81}  
Robinson E.L., Barker, E.S., Cochran, A.L., Cochran, W.D.,
Nather, R.E. 1981, \apj, 251, 611   

\bibitem[Rodgers et al. (1993)]{rod93}  
Rodgers, A.W., Roberts, W.H., \& Walker, I. 1993, \aj, 106, 591  

\bibitem[Rodgers \& Roberts (1994)]{rod94}  
Rodgers, A.W., \& Roberts, W.H. 1994, IAUC 6043

\bibitem[Sahnow et al. (2000)]{sah00}
Sahnow, D.J. et al. 2000, \apj, 538, L7 

\bibitem[Sarna et al. (1995)]{sar95} 
Sarna, M.J., Pych, W., \& Smith, R.C. 1995, IBVS No. 4165 

\bibitem[Schmid (1989)]{sch89}  
Schmid, H.M. 1989, \aap, 211, L31  

\bibitem[Schmid et al. (1999)]{sch99}  
Schmid, H.M. et al. 1999, \aap, 348, 950   

\bibitem[Schneider et al. (1981)]{sch81}  
Schneider, D.P., Young, P., Shectman, S.A. 1981, \apj, 245, 644  

\bibitem[Schoembs \& Vogt (1981)]{scho81}  
Schoembs, R., \&  Vogt, N. 1981, \aap, 97, 185 

\bibitem[Shafter (1983)]{sha83}  
Shafter, A. 1983, Ph.D. Thesis, UCLA  

\bibitem[Shafter \& Harkness (1986)]{sha86}  
Shafter, A. \& Harkness, R.P. 1986, \apj, 92, 685 

\bibitem[Shakun (1987)]{sha87}  
Shakun, L.I. 1987, Astronomicheskii Tsirkulyar, 1491, 7  

\bibitem[Shakura \& Sunyaev (1973)]{sha73} 
Shakura, N.I., \& Sunyaev, R.A., 1973, \aap, 24, 337                 

\bibitem[Shaviv \& Wehrse (1991)]{sha91}  
Shaviv, G. \& Wehrse, R. 1991, \aap, 251, 117  

\bibitem[Sheets et al. (2007)]{she07}  
Sheets, H.A., Thorstensen, J.R., Peters, C.M, Kapusta, A.B., Taylor, C.J., 
2007, \pasp, 119, 494  

\bibitem[Simon (2000)]{sim00}  
Simon, V. 2000, \aap, 354, 103  

\bibitem[Sion (1985)]{sio85}  
Sion, E.M. 1985, \apj, 292, 601  

\bibitem[Sion \& Urban (2002)]{sio02}  
Sion, E.M. \&  Urban, J. 2002, \apj, 572, 456  

\bibitem[Sion et al. (2012)]{sio12}   
Sion, E.M., Bond, H.E., Lindler, D., Godon, P., Wickramasinghe, D.,
Ferrario, L., \& Dupuis, J. 2012, \apj, 751, 66    

\bibitem[Sion et al. (2004)]{sio04a}  
Sion, E.M., Cheng, F., Godon, P., Urban, J.A., Szkody, P. 2004a, 
\aj, 128, 183 

\bibitem[Sion et al. (1996)]{sio96}  
Sion, E.M., Cheng, F.H., Huang, M., Hubeny, I., \& Szkody, P. 1996, 
\apj, 471, L41

\bibitem[Sion et al. (1995a)]{sio95a}  
Sion, E.M., Cheng, F., Long, K.S., Szkody, P., Huang, M., Gilliland, R., 
\& Hubeny, I. 1995a, \apj, 439, 957

\bibitem[Sion et al. (1997)]{sio97}  
Sion, E.M., Cheng, F.H., Sparks, W.M., Szkody, P., Huang, M., \& 
Hubeny, I. 1997, \apj, 480, L17 

\bibitem[Sion et al. (2008)]{sio08}  
Sion, E.M., G\"ansicke, B.T., Long, K.S., Szkody, P., Knigge, C.,
Hubeny, I., de Martino, D., Godon, P., 2008, ApJ, 681, 543  

\bibitem[Sion et al. (2007)]{sio07} 
Sion, E.M., Godon, P., Cheng, F., Szkody, P., 2007, \aj, 134, 886  

\bibitem[Sion et al. (2010)]{sio10} 
Sion, E.M., Godon, P., Myszka, J., Blair, W.P., 2010, \apj, 716, L157   

\bibitem[Sion et al. (1998)]{sio98}  
Sion, E.M., Schaeffer, K.G., Bond, H.E., Saffer, R.A.,
Cheng, F.H. 1998, \apj, 496, L29 

\bibitem[Sion et al. (1995b)]{sio95b}  
Sion, E.M., Szkody, P., Cheng, F.H., \& Huang, M. 1995b, \apj, 444, L97

\bibitem[Sion et al. (2001)]{sio01}  
Sion, E.M., Szkody, P., Cheng, F.H., G\"ansicke, B., LaDous, C., \& 
Hassal, B.J.M. 2001, \apj, 555, 834 

\bibitem[Sion et al. (2004a)]{sio04b}  
Sion, E.M., Winter, L., Urban, J.A., Tovmassian, G.H., Zharikov, S., 
G\"ansicke, B.T., Orio, M.  2004b, \aj, 128, 1795 

\bibitem[Skillman \& Patterson (1988)]{ski88}  
Skillman, D.R., \& Patterson, J. 1988, \aj, 96, 976   

\bibitem[Skillman et al. (1995)]{ski95}    
Skillman, D.R., Patterson, J., Thorstensen, J.R. 1995, \pasp, 107, 545  

\bibitem[Skopal (2005)]{sko05}  
Skopal, A. 2005, \aap, 440, 995  

\bibitem[Skopal et al. (2009)]{sko09}  
Skopal, A., Seker\'as, M., Gonz\'alez-Riestra, R., Viotti, R.F. 2009, \aap, 507, 1531 

\bibitem[Skopal et al. (2007)]{sko07}  
Skopal, A., Vanko, M., Pribulla, T., Chochol, D., Semkov, E., Wolf, M.,
\& Jones, A. 2007, Astron. Nachr., 328, 909  

\bibitem[Smith et al. (1996)]{smi96}  
Smith, V.V., Cunha, K., Jorissen, A., Boffin, H.M.J. 1996, \aap, 315, 179  

\bibitem[Smith et al. (1997)]{smi97}  
Smith, R.C., Sarna, M.J., Catalan, M.S., \& Jones, D.H.P. 1997, 
\mnras, 287, 271 

\bibitem[Spogli et al. (2003a)]{spo03a}  
Spogli, C., Fiorucci, M., Raimondo, G. 2003a, IBVS 5365, 1  

\bibitem[Spogli et al. (2003b)]{spo03b}  
Spogli, C., Fiorucci, M., Dolci, M., Raimondo, G. 2003b, IBVS 5474, 1  

\bibitem[Spogli et al. (2005)]{spo05}  
Spogli, C., et al. 2005, IBVS 5996, 1  

\bibitem[Sproats et al. (1996)]{spr96}  
Sproats, L.N., Howell, S.B., \& Mason, K.O. 1996, \mnras, 282, 1211

\bibitem[Stokes et al. (2000)]{sto00} 
Stokes, S.J., Evans, J.M., Bianchini, A., Canterna, R., 2000, AAS, 197.8505 

\bibitem[Stover (1981)]{sto81b} 
Stover, R. 1981, \apj, 249, 673  

\bibitem[Stover et al. (1981)]{sto81}  
Stover, R.J., Robinson, E.L., Nather, R.E. 1981, \apj, 248, 696 

\bibitem[Stump \& Sion (2001)]{stu01}  
Stump, M., \& Sion, E.M. 2001, \pasp, 113 1222  

\bibitem[Szkody \& Downes (1982)]{szk82}  
Szkody, P., \& Downes, R.A. 1982, \pasp, 94, 328  

\bibitem[Szkody \& Mattei (1984)]{szk84}  
Szkody, P., \& Mattei, J.A. 1984, \pasp, 96, 988  

\bibitem[Szkody et al. (1981)]{szk81}  
Szkody, P., Crosa, L., Bothun, G.D., Downes, R.A., \& Schommer, R.A. 
1981, \apj, 249, L61 

\bibitem[Szkody et al. (1988)]{szk88}  
Szkody, P., Downes, R., \& Mateo, M. 1988, \pasp, 100, 362 

\bibitem[Tappert et al. (1997)]{tap97}  
Tappert, C., Wargau, W., Hanuschik, R.W., \& Vogt, N. 1997, \aap, 327, 231

\bibitem[Taylor \& Thorstensen (1996)]{tay96} 
Taylor, C.M., \&  Thorstensen, J.R. 1996, PASP, 108, 894  

\bibitem[Thorstensen (1986)]{tho86}  
Thorstensen, J.R., 1986, \aj, 91, 940  

\bibitem[Tomov et al. (2000)]{tom00}  
Tomov, N., Tomova, M., Ivanova, A. 2000, \aap, 364, 557  

\bibitem[Tomov \& Tomova (2002)]{tom02}  
Tomov, N., Tomova, M. 2002, \aap, 388, 202  

\bibitem[Tovmassian et al. (2002)]{tov02}  
Tovmassian, G., Orion, M., Zharikov, S., Echevarria, J., 
Costero, R., \& Michel, R. 2002, in AIP Conf.Proc.637, Classical Nova 
Explosions, ed.M.Hernanz \& J.Jose (Melville, NY: AIP), 72 

\bibitem[Urban \& Sion (2006)]{urb06}  
Urban, J.A., Sion, E.M. 2006, \apj, 642, 1029   

\bibitem[Van Altena et al. (1995)]{vana95}  
Van Altena, W.F., Lee, J.T., Hoffleit, D.E. 1995, The General Catalog
of Trigonometric Parallaxes, 4th Edition, Yale University Observatory   

\bibitem[van Amerongen et al. (1987a)]{van87a}  
van Amerongen, S., Damen, E., Groot, M., Kraakman, H., \& 
van Paradijs, J. 1987a, \mnras, 225, 93

\bibitem[van Amerongen et al. (1987b)]{van87b}  
van Amerongen, S., Bovenschen, H., \& van Paradijs, J. 1987b, 
\mnras, 229, 245 

\bibitem[van Teeseling et al. (1995)]{van95}  
van Teeseling, A., Drake, J.J., Drew, J.E., Hoare, M.G., Verbunt, F.
1995, \aap, 300, 808  

\bibitem[Verbunt (1987)]{ver87}
Verbunt, F. 1987, \aap, 89, 223 

\bibitem[Viotti et al. (1984)]{vio84}  
Viotti, R., Altamore, A., Baratta, G.B., Cassatella, A., \& Friedjung, M. 1984, \apj, 283, 226  

\bibitem[Viotti et al. (1983)]{vio83}  
Viotti, R., Ricciardi, O., Giangrande, A., Ponz, D., Friedjung, M., 
Cassatella, A., Baratta, G.B., Altamore, A. 1983, \aap, 119, 285  

\bibitem[Vogt \& Bateson (1982)]{vog82} 
Vogt, N., and Bateson, F.M. 1982, \aaps, 48, 383  

\bibitem[Vogt \& Breysacher (1980)]{vog80}  
Vogt, N. \& Breysacher, J. 1980, \apj, 235, 945   

\bibitem[Vogt \& Semeniuk (1980)]{vog80s}  
Vogt, N. \&  Semeniuk, I. 1980, \aap, 89, 223  

\bibitem[Wade (1982)]{wad82}  
Wade, R. 1982, \aj, 87, 1558  

\bibitem[Wade (1988)]{wad88} 
Wade, R.A. 1988, \apj, 335, 394  

\bibitem[Wade \& Hubeny (1998)]{wad98} 
Wade, R.A., \& Hubeny, I. 1998, \apj, 509, 350 
 
\bibitem[Walter (2003)]{wal03}   
Walter, F. 2003, in the Future of Cool-Star Astrophysics: 12th Cambridge
Workshop on Cool Stars, Stellar Systems, and the Sun, eds. A. Brown, 
G.M. Harper, T.R. Ayres (University of Colorado), 14 

\bibitem[Walter (2004)]{wal04}   
Walter, F. 2004, A.N., 325, 241  

\bibitem[Wargau et al. (1983)]{war83}  
Wargau, W., Drechsel, H., Rahe, J., Bruch, A. 1983, \mnras, 204, 35P   

\bibitem[Wargau et al. (1984)]{war84}  
Wargau, W., Drechsel, H., Rahe, J., Bruch, A., Schombs, R. 1984, 
\apss, 99, 145   

\bibitem[Warner (1982)]{war82}  
Warner, B. 1982, Inf.Bul.Var.Stars, No.2175  

\bibitem[Warner (1985)]{war85}  
Warner, B., Odonoghue, D., Allen, S. 1985, \mnras, 212, 9  

\bibitem[Warner (1987)]{war87}  
Warner, B. 1987, \mnras, 227, 23   

\bibitem[Warner (1995)]{war95} 
Warner, B. 1995, Cataclysmic Variable Stars (Cambridge: Cambridge Univ.
Press) 

\bibitem[Watson et al. (2007a)]{wat07a}  
Watson, C.A., Steeghs, D., Shahbaz, T., Dhillon, V.S. 2007a, \mnras, 382, 1105 

\bibitem[Watson et al. (2007b)]{wat07b}  
Watson, C.A., Steeghs, D., Dhillon, V.S., Shahbaz, T. 2007b, Astron.Nachr., 328, 813 

\bibitem[Welsh et al. (2005)]{wel05}  
Welsh, W.F., Froning, C.S., Marsh, T.R., Robinson, E.L., Wood, J.H.,
2005, ASPC 330, 351   

\bibitem[Welsh et al. (2007)]{wel07}  
Welsh, W.F., Froning, C.S., Marsh, T.R., Reimer, T.W., Robinson, E.L.,
Wood, J.P.R. 2007, ASPC 362, 241  

\bibitem[Werner \& Rauch (1997)]{wer97}  
Werner, K., \& Rauch, R. 1997, \aap, 324, L25 

\bibitem[Wheatley (1998)]{whe98}  
Wheatley, P.J. 1998, \mnras, 297, 1145  

\bibitem[Wheatley et al. (1996)]{whe96}  
Wheatley, P.J., Verbunt, F., Belloni, T., Watson, M.G., Naylor, T., Ishida, M.,
Duck, S.R., Pfeffermann, E. 1996, \aap, 307, 137 

\bibitem[Williams (1983)]{will83}  
Williams, G., 1983, \apjs, 53, 523 

\bibitem[Williger et al. (1988)]{wil88}  
Williger, G., Berriman, G., Wade, R.A., Hassall, B.J.M. 1988, \apj, 333, 277 

\bibitem[Winter \& Sion (2001)]{win01}  
Winter, L. \&  Sion, E.M. 2001, BAAS, 33, 1399  

\bibitem[Winter \& Sion (2003)]{win03}  
Winter, L. \&  Sion, E.M. 2003, \apj, 582, 352  

\bibitem[Woods et al. (1992)]{woo92}  
Woods, J.A., Verbunt, F., Collier Cameron, A., Drew, J.E., \& Piters, A. 1992, \mnras, 255, 237  

\bibitem[Wood (1995)]{woo95} 
Wood, M.A. 1995, in White Dwarfs, Proceedings of the 9th
European Workshop on White Dwarfs, Lecture Notes in Physics,
Vol.442, eds. Detlev Koester \& Klaus Werner, Springer-Verlag, 
Berlin Heidelberg New York, p.41   

\bibitem[Young \& Schneider (1981)]{you81}  
Young, P., \&  Schneider, D.P. 1981, \apj, 247, 960  

\bibitem[Young \& Nelson (1972)]{you72}  
Young, A., \&  Nelson, B. 1972, \apj, 173, 653   

\bibitem[Young et al. (1983)]{you83}  
Young, A., Klimke, A., Africano, J.L., Quigley, R., Radick, R.R., 
van Buren, D. 1983, \apj, 267, 655  

\bibitem[Young et al. (2005)]{you05}  
Young, P.R., Dupree, A., Espey, B.R., Kenyon, S.J., Ake, T.B. 2005, \apj, 618, 891 

\bibitem[Zacharias et al. (2004)]{zac04}  
Zacharias, N., Urban, S.E., Zacharias, M.I., Wycoff, G.L., Hall, D.M.,
Monet, D.G., Rafferty, T.J. 2004, \aj, 127, 3043  

\bibitem[Zhang (1989)]{zha89}  
Zhang, E., 1989, Publications of the Beijing Astronomical Observatory, 13, 37 

\end{thebibliography}
\end{document}